\documentclass[12pt]{article}

 \usepackage{graphicx,color} 
\usepackage{amssymb,amsthm, amsmath, rotating}
\usepackage{amsbsy,amsfonts,amsgen}
\usepackage{subeqnarray}
\usepackage{fullpage}
\usepackage{subfigure}
\usepackage{authblk} 
\usepackage{chngcntr} 

\usepackage{lineno}
\usepackage{setspace}

\counterwithin{equation}{section} 

\usepackage{caption}
\DeclareCaptionType{mytype}[Table][List of mytype]  

\usepackage{graphics}
\usepackage{xcolor,graphicx}
\usepackage{epstopdf}
\graphicspath{{./}{../../figs/phosphate/slomp/}}

\newcommand{\lab}[1]{\label{#1}}
\newcommand{\ben}{\begin{enumerate}}
\newcommand{\een}{\end{enumerate}}
\newcommand{\be}{\begin{equation}}
\newcommand{\ee}{\end{equation}}
\newcommand{\bdm}{\begin{displaymath}}
\newcommand{\edm}{\end{displaymath}}
\newcommand{\bea}{\begin{eqnarray}}
\newcommand{\eea}{\end{eqnarray}}
\newcommand{\bean}{\begin{eqnarray*}}
\newcommand{\eean}{\end{eqnarray*}}
\newcommand{\bmat}[1]{\left(\begin{array}{#1}}
\newcommand{\emat}{\end{array}\right)}

\newcommand{\no}{\noindent}

\newcounter{question}

\newcommand{\reseteqnos}[1]{\renewcommand{\theequation}{#1.\arabic{equation}}
	\setcounter{equation}{0}}

\newcommand{\resetsectnos}[1]{\renewcommand{\thesection}{#1.\arabic{section}}
	\setcounter{section}{0}
	}

\newcommand{\bex}{\section*{Exercises}
\addcontentsline{toc}{section}{\protect\numberline{}{Exercises}}
\begin{list}%
{\arabic{chapter}.\arabic{question}}{\usecounter{question}}
}
\newcommand{\eex}{\end{list}}

\newcommand{\beax}[1]{\section*{Exercises}
\addcontentsline{toc}{section}{\protect\numberline{}{Exercises}}
\begin{list}%
{#1.\arabic{question}}{\usecounter{question}}
}
\newcommand{\eeax}{\end{list}}

\def\eps{\varepsilon}
\def\={&=&}

\def\dnon{\dfrac{}{}\nonumber\\}

\def\rmfor{\ \  {\rm for}\ \  }

\def\co2{CO$_2$}
\def\h2o{H$_2$O}
\def\Sio2{SiO$_2$}

\def\ch2o{CH$_2$O}

\def\12th{\tfrac{1}{12}}
\def\24th{\tfrac{1}{24}}
\def\48th{\tfrac{1}{48}}

\def\3.4ths{\tfrac{3}{4}}

\newcommand{\up}[1]{$^{{#1}}$}

\def\ie{i.\,e.,\ }
\def\eg{e.\,g.,\ }
\def\etal{{\it et al}.\ }

\newcommand{\de}{\delta}

\newcommand{\Tht}{\Theta}

\newcommand{\la}{\lambda}

\title{The development of deep-ocean anoxia in a comprehensive ocean phosphorus model}

\author[1]{J.\,G.\ Donohue}
\author[2]{B.\,J.\ Florio}
\author[1,3]{A.\,C.\ Fowler\thanks{Corresponding author: andrew.fowler@ul.ie}}
\affil[1]{MACSI,
University of Limerick, Limerick, Ireland}
\affil[2]{Department of Education, Western Australia, Australia}
\affil[3]{OCIAM,
University of Oxford,
Oxford, UK}

\date{}

\begin{document}

\maketitle

\begin{abstract}

We analyse a model of the phosphorus cycle in the ocean given by Slomp and Van Cappellen (2007, https://doi.org/10.5194/bg-4-155-2007). This model contains four distinct oceanic boxes and includes relevant parts of the water, carbon and oxygen cycles. We show that the model can essentially be solved analytically, and its behaviour completely understood without recourse to numerical methods. In particular, we show that, in the model, the carbon and phosphorus  concentrations in the different ocean reservoirs are all slaved to the concentration of soluble reactive phosphorus in the deep ocean, which relaxes to an equilibrium on a time scale of 180,000 y, and we show that the deep ocean is either oxic or anoxic, depending on a critical parameter which we can determine explicitly. Finally, we examine how the value of this critical parameter depends on the physical parameters contained in the model. The presented methodology is based on tools from applied mathematics and can be used to reduce the complexity of other large, biogeochemical models.

\end{abstract}

{\em Keywords: phosphorus cycle, mathematical model, ocean anoxia event, model reduction}. \\

\section{\lab{sec1}Introduction}
\reseteqnos{1}

There are two obvious reasons for wishing to study the phosphorus cycle in the world's oceans. The first is that it is intimately linked to variations in oxygen, carbon and other elements, both in the atmosphere and in the oceans,  and hence also to climate (Van Cappellen and Ingall 1996, Mackenzie \etal 2002). The phosphorus cycle is closely tied to the biological cycle, particularly in the oceans. While on land either phosphorus or nitrogen may be the limiting nutrient, in the ocean it is phosphorus that is believed to be the limiter on geological time scales. This is due to the population of algae (nitrogen fixers) which are able to source nitrogen from the atmosphere (Tyrell 1999).

The second reason is  that in order to fully understand the effect of anthropogenic alteration of the nitrogen and phosphorus cycle through the use of agricultural fertilisers, an understanding of the underlying processes and their time scales of operation is necessary, particularly in view of the impending phosphate crisis  (Abelson 1999, Cordell \etal 2009).

The phosphorus (or phosphate) cycle has been frequently described (Filipelli 2002, 2008, F\"ollmi 1996), but in order to assess and parameterise its effects in the geological past, it is necessary to describe the system using a mathematical model. A number of such models have been put forward (\eg Van Capellen and Ingall 1994, Anderson and Sarmiento 1995, Lenton and Watson 2000, Bergman \etal 2004, Tsandev \etal 2008, Ozaki \etal 2011), with various applications in mind.

One particular application of much recent interest has to do with the occurrence of `oceanic anoxia events' (OAEs), which have occurred in the geological past, particularly in the Jurassic and  Cretaceous periods (Schlanger and Jenkyns 1976, Jenkyns 2010). These events are marked in the marine sedimentary record by the occurrence of organically rich `black shales', and mark periods (of hundreds of thousands of years) during which the deep ocean became anoxic, thus promoting anaerobic digestion and the production of sulphides and other reduced substances.

It has become increasingly clear that OAEs are frequently associated with the formation of large igneous provinces (LIPs) (Turgeon and Creaser 2008, Sell \etal 2014, Percival \etal 2015), and that these may also be associated with increased weathering (Percival \etal 2016), as well as extinction episodes, which themselves might be due to increased  upwelling of anoxic water (Jarvis \etal 2008).

OAEs are also associated with severe changes in climate: warming occurs due to carbon change in the atmosphere, leading to enhanced precipitation and weathering, hence increased nutrient supply to the oceans, and consequent biomass blooms: this causes increased oxygen demand in the upper ocean, and this can lead to deep ocean anoxia (Jenkyns 2010).
Eutrophic conditions in the surface ocean may be further enhanced by redox-dependent release of phosphorus from anoxic sediments (Van Capellen and Ingall 1994).
In view of anthropogenic climate change, this raises the question as to whether ocean anoxia is a prospective consequence of present rates of atmospheric carbon increase (Watson 2016).
On the other hand, Niemeyer \etal (2017) suggest that the positive benthic P-release feedback may be mitigated by the configuration of the modern ocean, preventing a full-scale OAE.

It is clear that the mechanisms through which OAEs are sustained are controversial (Beil \etal 2020) with evidence often generated through the simulation of detailed numerical box models, for example those of Handoh and Lenton (2003), Slomp and Van Capellen (2007) and Wallmann \etal (2019). Thus, there is a need to enhance understanding of how these models produce a prediction, rather than allowing them to become black boxes (Maeda \etal 2021). Unfortunately, a common feature of such models is their inaccessibility; typically a large number of variables in a number of oceanic `boxes' describe the concentrations of various chemical components, and these are governed by differential equations which relate changes of the concentrations to reaction terms and inter-box fluxes. The complexity of the models is visible even in the opacity of their presentation, and their solution is inevitably obtained through numerical simulation. Because of this, it is difficult to interrogate the models and virtually impossible to unravel key mechanisms which control the dynamics.

The purpose of this paper is to present a methodology, based on tools of applied mathematics, which can be used to digest such complicated models, and reduce them to a form where their solutions can be obtained cheaply and simply, and the behaviour of the model can be specifically interpreted in terms of the prescribed parameters of the model.

In particular, we provide an exegesis of the model of Slomp and Van Cappellen (2007), which elaborated the model of Van Cappellen and Ingall (1996) to take account of the difference between continental shelves and the deep ocean. They were particularly interested in the effects of ocean mixing on phosphorus burial, and consequently on deep ocean anoxia. The numerical results from this model (henceforth called the Slomp model) indicate that oxygen concentration and mixing between boxes significantly affects the phosphorus cycle: in particular, they say: ``the simulations show that changes in oceanic circulation may induce marked shifts in primary productivity and burial of reactive phosphorus between the coastal and open ocean domains''. Our aim will be to provide explicit parametric interpretation of their results.

Our methods, while simple in concept, are sophisticated in practice. They are based on the ideas of non-dimensionalisation, scaling, and then asymptotic simplification. As is often the case, the simplifications arise because most of the describing equations act on a faster time scale than the slowest, and thus rate-controlling, equations. This allows us to achieve our goal. In the rest of the paper, the model is described and presented in section \ref{sec2}, and it is then non-dimensionalised in section \ref{sec2.1}. The resulting non-dimensional model is incorrectly scaled; we identify the reason for this, and correct the problem (by rescaling appropriately). The resulting asymptotic simplifications are described in \ref{sec2.2}, and lead to the result that all the ocean variables are slaved to the deep ocean soluble reactive phosphorus, which relaxes to an equilibrium on a time scale of 180,000 y.

In section \ref{sec3}, we show that the deep ocean oxygen and reduced substances concentrations can be determined analytically, and we show that there is a switch from an oxic deep ocean to an anoxic deep ocean at a critical value of one of the dimensionless parameters. In section \ref{sec4}, we endeavour to unravel the interpretation of our results in terms of the physical processes and parameters of the problem; this is the section where the mathematics-averse should go. Finally we offer our conclusions in section \ref{sec5}. We consign much of the algebraic debris to the appendix.

\section{\lab{sec2}The Slomp and van Capellen model}
\reseteqnos{2}

The Slomp model divides the  ocean into four distinct boxes: proximal coastal, distal coastal, surface ocean and deep ocean, having volumes $W_1$--$W_4$. Volume fluxes between the boxes are denoted by $WF_i$, $i=1,2,\ldots,7$. The boxes and fluxes are shown in figure \ref{fig1}.
As shown in table \ref{table6}, the fluxes corresponding to river input (${WF}_1$), ocean upwelling (${WF}_5$) and coastal upwelling (${WF}_6$) are defined empirically, via constants that we will refer to as ${Wk}_1$, ${Wk}_5$ and ${Wk}_6$ respectively.
Changes in circulation are modelled by multiplying the oceanic and coastal upwelling constants by the non-dimensional parameters $v_o$ and $v_c$, respectively.
The remaining four fluxes in the oceanic circulation system then arise by imposing conservation of stationary water volume within each box. The values assigned to the water-cycle parameters are listed in table \ref{table4} of the appendix.

\begin{figure}[t] %
 \centering
\includegraphics[width=\textwidth]{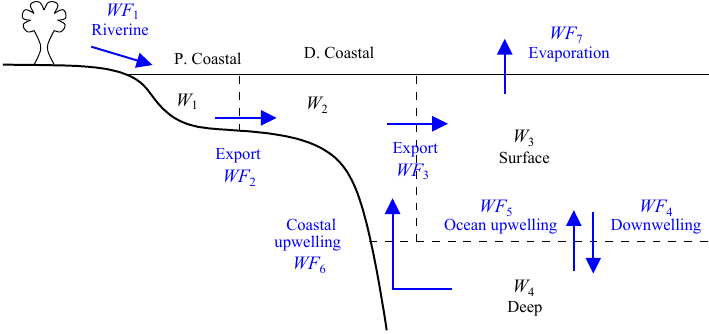}
 \caption{Modelling the oceanic water cycle as four distinct boxes with $W_1$--$W_4$ representing water volumes and $WF_i$, $i=1,2,\ldots,7$ representing volume fluxes between the boxes. The volume fluxes are defined in table \ref{table6}. }
\lab{fig1}
\end{figure}

\begin{table}[!t]
\begin{center}
\centering
\begin{tabular}{|l|l|l|l|l|}
\hline
Label  & Flux & Definition  \\
\hline
${WF}_1$ & River input & ${Wk}_1$   \\
${WF}_2$ & Proximal to distal & $={WF}_1$   \\
${WF}_3$ & Distal to surface & $={WF}_2 + {WF}_6$  \\
${WF}_4$ & Ocean downwelling & $={WF}_5 + {WF}_6$   \\
${WF}_5$ & Ocean upwelling & $v_o {Wk}_5$  \\
${WF}_6$ & Coastal upwelling & $v_c {Wk}_6$  \\
${WF}_7$ & Evaporation & $={WF}_1$  \\
\hline
\end{tabular}
\end{center}
\caption{Definition of water fluxes in the Slomp model. The values of the constants ${Wk}_1$, ${Wk}_5$, ${Wk}_6$, $v_o$ and $v_c$ are given in table \ref{table4} of the appendix.}
\lab{table6}
\end{table}

The model describes the quantities of phosphorus, carbon and oxygen in the different oceanic boxes. Phosphorus is assumed to be in one of three forms: reactive (SRP), organic particulate (POP), or authigenic calcium phosphate (fish hard parts). The quantities in each box are denoted by $S_i$ (SRP), $P_i$ (POP) and $F_i$ (fish P). (Here we deviate from Slomp and van Cappellen (2007), who allocate $P_i$, $i=1,2,\ldots,12$ to these variables.) The  phosphorus budgets are altered either by reactive processes within an oceanic box, or by travelling from one box to another.

The carbon cycle is a good deal simpler. It is described by modelling particulate organic carbon (POC) and is associated with living and detrital biomass. POC may grow within an oceanic box depending on phosphorus levels, and additionally there are inter-box fluxes. The concentration of POC in box $i$ is denoted by $C_i$.

The modelling of the oxygen system is assumed to be important only for the deep ocean, $W_4$. The surface-level boxes, $W_1$, $W_2$ and $W_3$, are assumed to be fully oxic as they are in communication with the atmosphere. As such, we only model deep ocean oxygen budget $G_4$, which changes in response to water-cycle fluxes, and also aerobic respiration within $W_4$. In an oxygen depleted system, reduced substances like sulphides can be removed from the system via burial, upwelling or by being oxidised. The concentration of these reduced substances is denoted by $R_4$, and is measured in oxygen equivalents.
Importantly, the rate of microbial respiration divides the consumption of deep ocean organic carbon into two components, one of which uses oxygen as the terminal electron acceptor, while the other represents the use of reduced substances; the split between the two is taken to depend on the deep ocean oxygen concentration.
The full description of the model is given by Slomp and van Cappellen (2007), although some of the finer detail is only accessible through their Matlab code.

In all, the Slomp model thus consists of eighteen first order differential equations for the variables $C_i$, $P_i$, $S_i$, $F_i$, $G_4$ and $R_4$. The quantities $X_i$, $X=C,P,F,S,G,R$ are budgets, \ie measured in moles, but we prefer to write them as concentrations, thus  $x_i = X_i/W_i$, where $W_i$ are the volumes of the boxes. The fluxes of the water, phosphorus, carbon and oxygen cycles are associated with a set of parameters denoted $Wk_i$, $Pk_i$, $Ck_i$ and $Ok_i$ respectively. Their description and values are given in tables \ref{table4} and \ref{table5} of the appendix. The conversion of moles to concentrations produces a transformed version of this parameter set which is described in tables \ref{table3part1} and \ref{table3part2} of the appendix. We find that the converted model takes the form \newpage
\bea\lab{2.1}
\dot{c}_1 &=& b_3 s_1 - b_4 c_1 ,\dnon
\dot{c}_2 &=& b_5 c_1 + b_6 s_2 - (b_{81} v + b_7) c_2,\dnon
\dot{c}_3 &=& b_8 s_3 + (b_{82} v + b_9) c_2 - b_{10} c_3,\dnon
\dot{c}_4 &=& \{1 - b_{85} \phi(g)\}(b_{83} v + b_{11}) c_2 + \{1 - b_{85} \phi(g)\} b_{12} s_3 - b_{13} c_4,\dnon
\dot{s}_1 &=& a_{53} + a_{14} p_1 + a_{15} f_1 - m_{71} s_1 ,\dnon
\dot{p}_1 &=& m_{72} s_1 - a_{18} p_1 ,\dnon
\dot{f}_1 &=& m_{73} s_1 - a_{20} f_1,\dnon
\dot{s}_2 &=& a_{21} s_1 + a_{22} p_2 + a_{23} f_2 - (a_{74} v + m_{74}) s_2 + v m_{54} s_4,\dnon
\dot{p}_2 &=& m_{75} s_2 + a_{26} p_1 - (a_{81} v + a_{27}) p_2,\dnon
\dot{f}_2 &=& m_{76} s_2 - a_{29} f_2,\dnon
\dot{s}_3 &=& a_{30} p_3 + m_{56} v s_4 + (a_{82} v + a_{31}) s_2 - (m_{32} v + b_{32}) s_3 ,\dnon
\dot{p}_3 &=& m_{77} s_3 + (a_{83} v + a_{34}) p_2 - (b_{84} v + b_{35}) c_2 - a_{36}p_3,\dnon
\dot{f}_3 &=& m_{78} s_3 - a_{38} f_3,\dnon
\dot{s}_4 &=& a_{39} p_4 + a_{40} f_4 + v m_{58} s_3 - v m_{58} s_4 - a_{59} \psi(g),\dnon
\dot{p}_4 &=& (b_{86} v + b_{41}) \left[ R_{CP}^{-1} - \frac{b_{85}}{ (C/P)_{\mbox{\tiny oxic}} } \chi(g) \right]c_2 + b_{42} \left[R_{CP}^{-1} - \frac{ b_{85}}{ (C/P)_{\mbox{\tiny oxic}} } \chi(g) \right] s_3 - a_{43} p_4,\dnon
\dot{f}_4 &=& a_{44} f_3 - a_{45} f_4,\dnon
\dot{g} &=& v m_0 (g_s - g) - b_1 \frac{c_4 g}{K_m + g} - k_{\text{redox}} \Theta(r,g),\dnon
\dot{r} &=& b_{1} c_4 \left(1 - \frac{ g}{K_m + g} \right) - k_{\text{prec}} \theta(r) - k_{\text{redox}} \Theta(r, g),\eea
where we have written $g_4=g$ and $r_4=r$ as they have no counterparts in the other boxes.
The coefficients $a_i$ and $b_i$ are positive constants related to the phosphorus and carbon cycle respectively, whereas the coefficients $m_i$ are positive constants that do not fit neatly into either of the former categories. Their values are given in tables \ref{table3part1} and \ref{table3part2} of the appendix.
Furthermore, $R_{CP}$ is the Redfield ratio of carbon to phosphorus with $(C/P)_{\mbox{\tiny oxic}}$ and $(C/P)_{\mbox{\tiny anoxic}}$ the ratios of carbon to phosphorus for sedimentary organic matter buried under oxic and anoxic conditions, respectively.
Finally, $k_{\rm redox}$ controls the reduced-substances reoxidation rate, $k_{\rm prec}$ controls the removal of reduced substances via precipitation, $K_m$ is a Monod constant, $g_s$ is the fully oxic surface concentration and $v$ is a dimensionless mixing parameter. The functions in \eqref{2.1} are defined by

\bea\lab{2.2}
\theta(r) &=&
                \left[ \dfrac{ r}{C_{RP}}  - 1 \right]_+, \dnon
\Theta(r, g) &=& 10^{-6}rg,\dnon
\psi(g) &=& \begin{cases}
             \dfrac{g}{g_0} \rmfor g < g_0,\\
             1 \rmfor g \geq g_0,
            \end{cases}\dnon
\chi (g) &=& \begin{cases}
           0.75 + \dfrac{0.25 g}{g_0} \rmfor g < g_0,\\
             1 \rmfor g \geq g_0,
            \end{cases}\dnon
 \phi(g) &=& \begin{cases}
             \left(0.75 + \dfrac{0.25 g}{g_0} \right) \left[ \dfrac{g}{g_0} + \dfrac{ (C/P)_{\mbox{\tiny oxic}} }{ (C/P)_{\mbox{\tiny anoxic}} } \left(1 - \dfrac{g}{g_0} \right) \right]^{-1} \rmfor g < g_0,\\
             1 \rmfor g \geq g_0,
            \end{cases}
\eea
where $g_0$ is a deep oxygen threshold, $[x]_+=\max(x,0)$, and we have assumed that $r>0$ and $g>0$ in the definition of $\Tht$. These functions correspond to the flux terms given in equations (3)--(7) of Slomp and van Capellen (2007). However, for $\chi$, $\psi$ and $\phi$, only the definitions corresponding to $g<g_0$ are reported in the article proper.
Note that the factor of $0.25$ in the definitions of $\chi$ and $\phi$ arises from the assumption that anoxia may reduce the burial flux of $p_4$ by up to 25\%.
Finally, we note that before the rate law (describing reoxidation of reduced substances) given in Slomp and van Capellen's equation (4) can be applied, we must first convert $r$ and $g$ from units of moles m$^{-3}$ to units of moles l\up{-1}. This leads to the factor of $10^{-6}$ in \eqref{2.2}.

\subsection{\lab{sec2.1}Non-dimensionalisation of the model}

Our procedure for simplifying the model begins by non-dimensionalising the system. Numerically, using the parameter values estimated by Slomp and van Capellen (2007), the solution approaches a steady state. Our aim is to
 scale the system so that the scaled concentration variables are $O(1)$ at this steady state.
We first note that in Slomp and van Capellen’s model, the mixing parameter $v$ was taken to be $=1$ for a well-mixed ocean, but was lower for poorly mixed anoxic oceans, with values $v\sim v_a=0.1$. We use this anoxic value in our choice of scales below, but because later in section 3.2 we also consider the case of a well-mixed ocean, it is useful to retain the dimensionless parameter
\be\lab{newvv}\nu=\frac{v}{v_a}\ee
in order to facilitate the possibility of adjusting the mixing parameter in a straightforward manner.

Next, we ensure that the scaled concentrations are $O(1)$ at the steady state by identifying the largest terms on the right hand sides of system \eqref{2.1} and balancing them. For example, let $s_1 = [s_1] \bar{s}_1$, where $[s_1]$ denotes the scale, and $\bar{s}_1$ is the new dimensionless variable. For some equations, the scaling argument is straightforward; consider for example \eqref{2.1}$_1$, which becomes
\begin{equation}\lab{2.3}
 \frac{[c_1]}{[t]}\dot{\bar{c}}_1 = b_3 [s_1] \bar{s}_1 - b_4 [c_1] \bar{c}_1,
\end{equation}
where $[t]$ is the chosen time scale.
A balance of the two terms on the right hand side gives
\begin{equation}\lab{2.4}
 0 = b_3[s_1] - b_4 [c_1],
\end{equation}
which relates the scales of $[s_1]$ and $[c_1]$. For equations with more than two terms, the results of a numerical simulation are used to infer the largest two terms. One must be careful in some situations where a cyclic definition of scales is found. For example, consider \eqref{2.1}$_5$ and \eqref{2.1}$_6$, where taking the largest two terms gives
\bea\lab{2.5}
0 &=& a_{14} [p_1] - m_{71} [s_1 ],\dnon
0 &=& m_{72} [s_1] - a_{18} [p_1],
\eea
for which the only solution is $[p_1] = [s_1] = 0$. To resolve this conundrum, we consider also the next largest terms. In this particular case, it is the constant riverine input, giving
\bea\lab{2.6}
0 &=& a_{53} + a_{14} [p_1] - m_{71} [s_1 ],\dnon
0 &=& m_{72} [s_1] - a_{18} [p_1],
\eea
which has a non-trivial solution. Physically, this occurs because a large amount of phosphorus is cycled between SRP and POP phases compared to the net input and output. Two further instances of this cyclicity occur in the choice of scales for $s_2$ and $s_3$. It is perhaps easier to see how scales are chosen by restricting ourselves to the most obvious balances. These are
\bea\lab{2.6.1}&&r\sim C_{RP},\quad 10^{-6}rg=\Tht\sim\frac{  m_0 v_a g_s}{k_{\text{redox}}},\dnon
&&s_1\sim\frac{b_4c_1}{b_3}\sim\frac{a_{18}p_1}{m_{72}}\sim\frac{a_{20}f_1}{m_{73}},\dnon
&&s_2\sim\frac{b_7 c_2}{b_6}\sim\frac{a_{29}f_2}{m_{76}}\sim\frac{a_{27}p_2}{m_{75}},\dnon
&&s_3\sim\frac{b_{10}c_3}{b_8}\sim\frac{b_{13}c_4}{b_{12}}\sim\frac{a_{36}p_3}{m_{77}}\sim\frac{a_{38}f_3}{m_{78}}\sim\frac{a_{43}R_{CP}p_4}{b_{42}},\dnon
&&s_4\sim\frac{a_{39}p_4}{ m_{58} v_a },\quad f_4\sim\frac{a_{44}f_3}{a_{45}},\quad t\sim\frac{1}{  m_{58} v_a }\sim  3,\!000 \mbox{ y},\eea
with  the time scale being chosen as the longest time scale of any of the equations, which  leads to the consequence that all of the time derivatives (bar that of the slowest equation) will be multiplied by parameters less than one (and in fact much less than one). The values associated with these scales are given in \eqref{B4} of the appendix. The resulting scaled system is given by \newpage
\bea\lab{2.7}
\varepsilon_{28} \dot{c}_1 &=& s_1 - c_1,\dnon%
\varepsilon_{29} \dot{c}_2 &=& \delta_{4} c_1 + s_2-\varepsilon_{103} \nu c_2-c_2,\dnon%
\varepsilon_{30} \dot{c}_3 &=& s_3+\varepsilon_{112} \nu c_2 +\varepsilon_{13} c_2 -c_3, \dnon%
\varepsilon_{31} \dot{c}_4 &=&  (\varepsilon_{14}+ \varepsilon_{113} \nu ) c_2  - (\varepsilon_{15} +  \varepsilon_{105} \nu)c_2 \phi(g)    +s_3    - \varepsilon_{16} s_3 \phi(g) -c_4,\dnon%
\varepsilon_{20} \dot{s}_1 &=& \lambda_{1}+ \lambda_{2} p_1  + \varepsilon_{1} f_1  -s_1, \dnon%
\varepsilon_{23} \dot{p}_1 &=& s_1 - p_1,\dnon%
\varepsilon_{27} \dot{f}_1 &=& s_1 - f_1,\dnon%
\varepsilon_{21} \dot{s}_2 &=& \varepsilon_{2} s_1   +\lambda_{3} p_2   +\varepsilon_{3} f_2   +\varepsilon_{4} \nu s_4     -\varepsilon_{101} \nu s_2   -s_2, \dnon
\varepsilon_{24} \dot{p}_2 &=& s_2 +\delta_{5} p_1  - \varepsilon_{102} \nu p_2 -p_2,\dnon%
\varepsilon_{34} \dot{f}_2 &=& s_2 - f_2,\dnon%
\varepsilon_{22} \dot{s}_3 &=& \lambda_{4} p_3+\delta_{1} \nu s_4   +\delta_{2} s_2+\lambda_{20} \nu s_2    -\varepsilon_{106} \nu s_3 -s_3, \dnon
\varepsilon_{25} \dot{p}_3 &=& s_3   + \varepsilon_{8} p_2  + \varepsilon_{110} \nu p_2      -\varepsilon_{9} c_2  - \varepsilon_{111} \nu c_2     -p_3,\dnon%
\varepsilon_{35} \dot{f}_3 &=& s_3 - f_3,\dnon%
                 \dot{s}_4 &=& p_4 +\varepsilon_{6} f_4  + \varepsilon_{107} \nu s_3  -s_4 \nu        - \lambda_{5} \psi(g) ,\dnon%
\varepsilon_{26} \dot{p}_4 &=& (\varepsilon_{33} + \varepsilon_{99} \nu) c_2 - (\varepsilon_{10} + \varepsilon_{104} \nu)c_2 \chi(g)    +s_3          - \varepsilon_{11} s_3 \chi(g)  -p_4,\dnon%
\varepsilon_{36} \dot{f}_4 &=& f_3 - f_4,\dnon%
\varepsilon_{32} \dot{g} &=& \nu(1 - \varepsilon_{19} g)    -  \varepsilon_{39} \frac{c_4 g}{\lambda_{11} + g}   - r g,\dnon
\varepsilon_{37} \dot{r} &=& \varepsilon_{38} c_4 \left( 1-\frac{g}{\lambda_{11}+g} \right)  - [r - 1]_+  - \delta_{3} r g,
\eea
where we have omitted the overbars for convenience.
The functions in these equations are defined by
\bea\lab{2.8}
 \phi(g) &=& \begin{cases}
		  \left(0.75 + 0.25\dfrac{g}{g_0^*}\right)\left\{\dfrac{g}{g_0^*} + \dfrac{ (C/P)_{\mbox{\tiny oxic}} }{ (C/P)_{\mbox{\tiny anoxic}} }\left(1-\dfrac{g}{g_0^*}\right)\right\}^{-1} \rmfor g < g_0^*,\\
		  1 \rmfor g \geq g_0^*,
               \end{cases}\dnon
 \psi(g) &=&  \begin{cases}
		    \dfrac{g}{g_0^*} \rmfor g < g_0^*,\\
		    1 \rmfor g \geq g_0^*,
                \end{cases}\dnon
 \chi(g) &=&  \begin{cases}
		    0.75 + 0.25\dfrac{g}{g_0^*} \rmfor g < g_0^*,\\
		    1 \rmfor g \geq g_0^*,
                \end{cases}
\eea
where  $g_0^* = g_0/[g]$.

The dimensionless coefficients are defined in \eqref{B1}--\eqref{B3} in the appendix. They are divided into three sets; parameters denoted $\la_i$ are of $O(1)$; parameters denoted $\de_i$ are small $\sim 0.1$, but not very small; and parameters $\eps_i$ are `very small', in practice $<0.1$. There is some fuzziness at the crossover, for example the parameters $\eps_6$, $\eps_8$, $\eps_{13}$,  $\eps_{14}$, $\eps_{33}$, $\eps_{99}$, $\eps_{107}$,  $\eps_{110}$, $\eps_{112}$ and $\eps_{113}$ could all have been taken as $\de$s.

The scales in  \eqref{2.6.1} give fifteen of the eighteen scales necessary, and it can be seen that of the equations, no precise balance has been applied in the equations for $s_1$, $s_2$ and $s_3$. As explained above, the scale for $s_1$ is chosen by solving \eqref{2.6}; this is equivalent to {\it choosing}
\be\lab{2.8.1}\la_2=1-\la_1.\ee
In a similar manner, the scales for $s_2$ and $s_3$ are chosen by defining
\bea\lab{2.8.2}\la_3\=1-\eps_2,\dnon
\la_4\=1-\de_2.\eea
This then completes the choice of scaling of the model. To determine if the scaling is appropriate for a poorly-mixed ocean, we now compute the dimensionless steady state solution with $\nu=1$; denoting these values with an asterisk, these are found to be
\begin{eqnarray} \lab{2.9}
 {g^*} = 0.55, &\qquad& c_1^* = 1.04, \dnon {c_2^*} = 26.61, &\qquad& c_3^* = 783.65, \dnon
 {c_4^*} = 774.87, &\qquad& s_1^* = 1.04, \dnon
 {p_1^*} = 1.04, &\qquad& f_1^* = 1.04, \dnon
 {s_2^*} = 26.6, &\qquad& p_2^* = 26.61, \dnon
 {f_2^*} = 26.6, &\qquad& s_3^* = 779.88, \dnon
 {p_3^*} = 783.7,  &\qquad& f_3^* = 779.88, \dnon
 {s_4^*} = 906.27, &\qquad& p_4^* = 782.81, \dnon
 {f_4^*} = 779.88, &\qquad& r^* = 1.14.
\end{eqnarray}
We might expect that the steady state values would be $O(1)$, but clearly this is not the case. Inspecting the sixteen carbon and phosphorus variables, it seems that there is a magnifying factor of about $30$ from box 1 to the corresponding box 2 variable, and then $30$ from box 2 to the corresponding box 3 variable. There is some subtle effect here, which needs to be elucidated. There are two key scales: $[s_2]$ and $[s_3]$; every other scale can be related back to these. We focus on the steady state solutions of the differential equations for $s_2$ and $s_3$. Substituting in the other variables and neglecting small terms, we can deduce
\bea\lab{2.10}
(\varepsilon_{101} \nu+\eps_{2}-\varepsilon_{3})s_2       -\varepsilon_{4} s_3 = (\lambda_{3} \delta_{5}+\varepsilon_{2}) s_1             -\lambda_{5} \varepsilon_{4} \psi(g),\ \  \quad &&\dnon
-((\lambda_{4} \varepsilon_{110}+\lambda_{20}) \nu+\delta_{2} ) s_2     + (\varepsilon_{106} \nu+\de_2 -\delta_{1}) s_3       = \lambda_{4} \delta_{5} \varepsilon_{110} s_1 \nu     -\delta_{1} \lambda_{5} \psi(g).&&
\eea
The coefficients of $s_2$ and $s_3$ on the left hand side of these equations form a $2\times2$ matrix which has a small determinant ($\approx 0.0007$) when $\nu=1$. This explains why the system is sensitive to inaccuracies. When $\nu=1$, the values of the diagonals of this matrix are $\varepsilon_{101} +\eps_{2}-\varepsilon_{3} \approx 0.029$ and $\varepsilon_{106} +\de_2 -\delta_{1} \approx 0.034$. In order to accommodate the fact that these numbers are  very small, it is appropriate to rescale the variables. We do this by defining rescaling parameters
\bea\lab{2.11}
 [\bar{s}_2] &=& \frac{1}{ \varepsilon_{101} +\eps_{2}-\varepsilon_{3} },\dnon
 {[\bar{s}_3]} &=& \frac{1}{( \varepsilon_{101} +\eps_{2}-\varepsilon_{3} )( \varepsilon_{106} +\de_2 -\delta_{1} )}.
\eea
Thus, from the original dimensionless variables, we now define  $\bar{s_2} = [\bar{s}_2] \hat{s}_2$, $\bar{s_3} = [\bar{s}_3] \hat{s}_3$, and from these we can deduce the rescaling of all the other variables  other than $r$, $g$ and those in box 1, which are unaltered, just as in \eqref{2.6.1}. The rescaled system is now found to be (in terms of the hatted variables, but again we drop the hats)
\bea\lab{2.13}
\varepsilon_{28} \dot{c}_1 &=& s_1 - c_1,\dnon%
\varepsilon_{29} \dot{c}_2 &=& \varepsilon_{40} c_1 + s_2-\varepsilon_{103} \nu c_2-c_2,\dnon%
\varepsilon_{30} \dot{c}_3 &=& s_3+\varepsilon_{123} \nu c_2 +\varepsilon_{41} c_2 -c_3, \dnon%
\varepsilon_{31} \dot{c}_4 &=&  (\varepsilon_{42}+ \varepsilon_{124} \nu ) c_2  - (\varepsilon_{43} +  \varepsilon_{126} \nu)c_2 \phi(g)    +s_3    -\varepsilon_{16} s_3 \phi(g) -c_4,\dnon%
\varepsilon_{20} \dot{s}_1 &=& \lambda_{1}+ \lambda_{2} p_1  + \varepsilon_{1} f_1  -s_1, \dnon%
\varepsilon_{23} \dot{p}_1 &=& s_1 - p_1,\dnon%
\varepsilon_{27} \dot{f}_1 &=& s_1 - f_1,\dnon%
\varepsilon_{21} \dot{s}_2 &=& \varepsilon_{44} s_1   +\lambda_{3} p_2   +\varepsilon_{3} f_2   +\varepsilon_{45} \nu s_4     -\varepsilon_{101} \nu s_2   -s_2, \dnon
\varepsilon_{24} \dot{p}_2 &=& s_2 + \varepsilon_{46} p_1  - \varepsilon_{102} \nu p_2 -p_2,\dnon%
\varepsilon_{34} \dot{f}_2 &=& s_2 - f_2,\dnon%
\varepsilon_{22} \dot{s}_3 &=& \lambda_{4} p_3+\delta_{1} \nu s_4   +\varepsilon_{47} s_2+ \varepsilon_{127} \nu s_2    -\varepsilon_{106} \nu s_3 -s_3, \dnon
\varepsilon_{25} \dot{p}_3 &=& s_3   + \varepsilon_{48} p_2  + \varepsilon_{121} \nu p_2      -\varepsilon_{49} c_2  - \varepsilon_{122} \nu c_2     -p_3,\dnon%
\varepsilon_{35} \dot{f}_3 &=& s_3 - f_3,\dnon%
                 \dot{s}_4 &=& p_4 +\varepsilon_{6} f_4  + \varepsilon_{107} \nu s_3  -s_4 \nu  - \varepsilon_{50} \psi(g) ,\dnon%
\varepsilon_{26} \dot{p}_4 &=& (\varepsilon_{51} + \varepsilon_{120} \nu) c_2 - (\varepsilon_{52} + \varepsilon_{125} \nu)c_2 \chi(g)    +s_3   -\varepsilon_{11} s_3 \chi(g)  -p_4,\dnon%
\varepsilon_{36} \dot{f}_4 &=& f_3 - f_4,\dnon%
\varepsilon_{32} \dot{g} &=& \nu(1 - \varepsilon_{19} g)    - \lambda_{6} \frac{c_4 g}{\lambda_{11} + g}   - r g,\dnon
\varepsilon_{37} \dot{r} &=& \delta_{25} c_4 \left( 1-\frac{g}{\lambda_{11}+g} \right)  - [r - 1]_+  - \delta_{3} r g.
\eea
The new dimensionless coefficients are defined in  \eqref{B5}. With $\nu=1$, the steady-state solution in these rescaled variables is given by
\begin{eqnarray}\lab{2.14}
 {g^*} = 0.551, &\qquad& c_1^* = 1.04,  \dnon
 {c_2^*} = 0.775, &\qquad& c_3^* = 0.786, \dnon
 {c_4^*} = 0.777, &\qquad& s_1^* = 1.04,  \dnon
 {p_1^*} = 1.04, &\qquad& f_1^* = 1.04,   \dnon
 {s_2^*} = 0.775, &\qquad& p_2^* = 0.775, \dnon
 {f_2^*} = 0.775, &\qquad& s_3^* = 0.782, \dnon
 {p_3^*} = 0.786, &\qquad& f_3^* = 0.782, \dnon
 {s_4^*} = 0.908, &\qquad& p_4^* = 0.785, \dnon
 {f_4^*} = 0.782, &\qquad& r^* = 1.14.
\end{eqnarray}
As they are now all $O(1)$, it shows that the current scaling is adequate for a poorly-mixed ocean.

\section{Model reduction}
\reseteqnos{3}

In this section, we study the dynamics of the scaled Slomp model. It is important to note that we will assume that the Slomp model has been well parameterised. Specifically, we will not allow for the possibility that their estimates of the system parameters differ from the `true' values by an order of magnitude or more. On this basis, a number of simplifications to the model can be made, as we will see in sections \ref{sec2.2} and \ref{sec3}.
It is also important to note that, in the analysis that follows, we will assume that the (dimensionless) initial conditions are $O(1)$ or equivalently that all variables are within an order of magnitude of their equilibrium values, as given by (\ref{2.14}). Through numerical investigation, it is apparent that this is the only stable steady-state solution associated with these parameter values. However, it is, of course, possible that there are additional steady-state solutions at other parameter values, as similar box models have been shown to exhibit bistable equilibria (Goldblatt \etal 2006) and sustained oscillations (Handoh and Lenton 2003, Wallmann \etal 2019). If such dynamics were to occur in the Slomp model, they could be traced back to one or more of the five nonlinear equations in \eqref{2.13}. Thus, our approach does not rule out the detection of more complex dynamics, though a search is not a focal part of the analysis.

\subsection{\lab{sec2.2}Simplification of the carbon-phosphorus model}

Inspecting \eqref{2.13}, it is clear that on a rapid time scale ($\sim\eps_i$), $s_1,c_1,p_1,f_1\to 1$. Similarly, in box 2, we rapidly have $c_2\approx s_2\approx p_2\approx f_2$, but the degeneracy between the $s_2$ and $p_2$ equations leaves their value indeterminate. As is usual in this situation, the missing information is obtained by eliminating the large term; we add the $s_2$ and $p_2$ equations, and this leads to (bearing in mind the box 1 and box 2 equalities)
\be\lab{2.15}(\eps_{21}+\eps_{24})\dot{s}_2=\eps_{44}+\eps_{46}+\eps_{45}\nu s_4-(\eps_2-\eps_3+\nu(\eps_{101}+\eps_{102}))s_2,\ee
suggesting a slower evolution of the box 2 variables.
Similarly, the box 3 concentrations all rapidly equilibrate, but there is degeneracy in the $s_3$ and $p_3$ equations, and adding these yields
\be\lab{2.16}(\eps_{22}+\eps_{25})\dot{s}_3=\de_1 \nu s_4+(\eps_{47}+\eps_{48}-\eps_{49} + (\eps_{121}+\eps_{127}-\eps_{122})\nu )s_2-(\de_2+\eps_{106}\nu)s_3.\ee
Finally,  the box 4 variables $f_4,c_4,p_4\to s_3$ rapidly, and thus the slow $s_4$ equation is
\be\lab{2.16.1}\dot{s}_4\approx (\la_9+\eps_{107} \nu)s_3-\nu s_4,\ee
where
\be\lab{2.16.2}\la_9=1+\eps_6 \approx 1.073.\ee
Thus, we can write the  $s_2$ and $s_3$ equations in the form
\bea\lab{2.17}\eps_{55} \dot{s}_2 \=\ \de_6 + \nu s_4-(\la_{13}+ \la_{14}\nu )s_2,\dnon
\eps_{56} \dot{s}_3 \=\  \nu s_4+(\de_7 + \de_8 \nu )s_2   - (\la_{15}  + \de_9 \nu )s_3,\eea
where
\bea\lab{2.18}&&\eps_{55}=\frac{\eps_{21}+\eps_{24}}{\eps_{45}}\approx 0.02,\quad \eps_{56}=\frac{\eps_{22}+\eps_{25}}{\de_1}\approx 0.68\times 10^{-2},\dnon
&&\de_6=\frac{\eps_{44}+\eps_{46}}{\eps_{45}}\approx 0.197,\quad \de_{7}=\frac{\eps_{47}+\eps_{48}-\eps_{49}}{\de_1}\approx 0.061, \dnon
&&\de_8=\frac{\eps_{121}+\eps_{127}-\eps_{122}}{\de_{1}}\approx 0.062, \quad \de_9= \frac{\eps_{106}}{\de_1}\approx 0.097  ,\dnon
&&\la_{13}=\frac{\eps_2-\eps_3}{\eps_{45}}\approx 0.818,\quad \la_{14}=\frac{\eps_{101}+\eps_{102}}{\eps_{45}}\approx 0.619,\dnon
&&\la_{15}=\frac{\de_2 }{\de_1}\approx 1.185.\eea
We have broken our rule about the size of $\de$s and $\eps$s, but it is necessary to retain the apparently small terms in $\de_7$, $\de_8$ and $\de_9$. Evidently the $s_2$ and $s_3$ equations are still relatively fast, and clearly both of them relax to an equilibrium approximately given by
\be\lab{2.19}s_2\approx \frac{\de_6+ s_4 \nu}{\la_{13} + \la_{14}\nu },\quad s_3\approx\frac{ ((\de_8+\la_{14}) \nu + \de_7+\la_{13} ) \nu s_4 + \de_6(\de_7+\de_8\nu ) }{ (\la_{15}+\de_{9} \nu ) (\la_{13}+\la_{14} \nu ) },\ee
following which $s_4$ relaxes to its equilibrium
\be\lab{2.20}s_4\approx \frac{ (\eps_{107} \nu+\la_{9}) \de_6 (\de_8 \nu+\de_7)}{ (\de_9 \la_{14}-(\de_8+\la_{14}) \eps_{107} ) \nu^3  +\eps_{57} \nu^2+( \la_{13} (\la_{15}-\la_9) - \la_{9} \de_7  ) \nu } \approx 0.943 \ee
with
\begin{equation} \eps_{57}= (\la_{15}-\la_9) \la_{14} -\eps_{107} (\de_7+\la_{13}) - \de_8 \la_{9}+\de_9 \la_{13}   \approx 4.04\times 10^{-3} \label{eps57defn}. \end{equation}
This relaxation occurs on a time scale
\be\lab{2.21}t\sim  \frac{ (\de_9 \nu+\la_{15})  (\la_{13} +\la_{14}\nu )}{ (\de_9 \la_{14}-(\de_8+\la_{14}) \eps_{107} ) \nu^3  +\eps_{57} \nu^2+( \la_{13} (\la_{15}-\la_9) - \la_{9} \de_7  ) \nu } \approx  61.6,\ee
corresponding to 180,000 y, much longer than our original time scale. Numerical verification of these analytical estimates is given in figure \ref{phostime} where the four SRP variables are used as exemplars.

    \begin{figure}
\centering
\includegraphics[width=\textwidth]{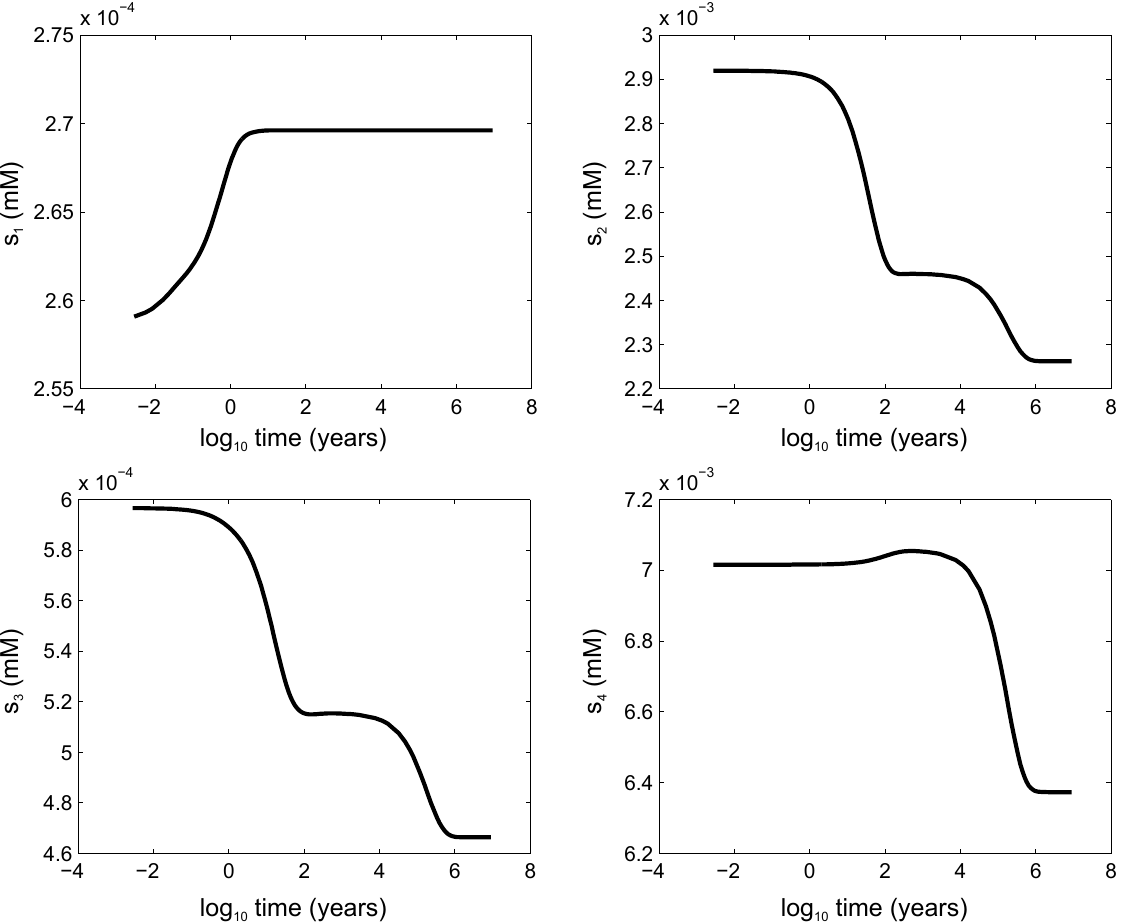}

\caption{ Equilibration of soluble reactive phosphorus concentrations in each oceanic box. All variables are presented in dimensional form and time has been logarithmically transformed in order to clearly illustrate the various time scales of interest. We have set the mixing parameter $\nu=1$ to represent a poorly-mixed ocean. The initial value of each of the eighteen system variables was set to be $1$ and \eqref{2.13} was solved numerically. 
}
\lab{phostime}
\end{figure}

\subsection{\lab{sec3}Oxygen dynamics}

Although the equations for $p_4$, $s_4$ and $c_4$ are coupled to $g$ in \eqref{2.13}, the coupling is weak and can be ignored, so that the carbon-phosphorus part of the model operates independently from oxygen and reduced substances in the deep ocean.
The model equations for $r$ and $g$ are given by the last pair in \eqref{2.13}, and depend on the carbon and phosphorus equations only through $c_4\approx s_3$, which is given by \eqref{2.19}, and varies on a slow time scale. Thus
\bea\lab{3.1}
 \varepsilon_{32} \dot{g} &=& \nu(1 - \varepsilon_{19} g) - \lambda_{6} \frac{s_3 g}{\lambda_{11} + g} - r g,\dnon
 \varepsilon_{37} \dot{r} &=& \de_{25} s_3 \left(1 - \frac{g}{\lambda_{11} + g}\right) - [r - 1]_+  -\delta_3 r g.
\eea
Now $\eps_{32}\sim 10^{-4}$ whereas $\eps_{37}\sim 10^{-2}$ and therefore the $g$ equation relaxes first to an equilibrium in which
\be\lab{3.2}r+\eps_{19}\nu=\frac{\nu}{g}-\frac{\la_6s_3}{\la_{11}+g}.
\ee
This defines $g$ as a decreasing function  $G(r)$, with $G(0)$ being finite or very large (as $\eps_{19} \sim 10^{-4}$) depending on whether $\la_6s_3>\nu$ or $<\nu$ respectively; figure \ref{fig3new} shows two typical examples, one with $s_3=0.782$ (corresponding to the steady state in \eqref{2.14}) and one using a much smaller value of $s_3$.

\begin{figure}
\begin{center}
\includegraphics[width=0.6\textwidth]{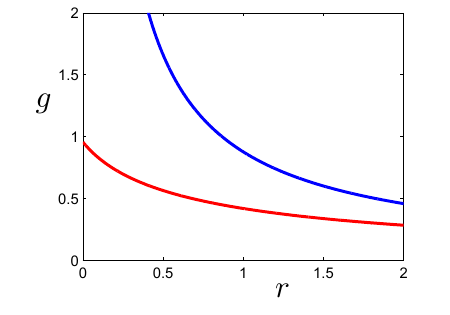}
\caption{The approximate slow manifold $g=G(r)$, or $g$ nullcline, \eqref{3.2}, using parameter values from the Slomp model and $\nu$ set to $1$ to represent a poorly-mixed ocean. In the lower curve, $s_3=0.782$ whereas in the upper curve, $s_3=0.1$. The variables $g$ and $r$ are dimensionless.}
\lab{fig3new}
\end{center}
\end{figure}

Following the relaxation of $g$ to its quasi-equilibrium $G(r)$, $r$ evolves (still relatively rapidly) via the $\varepsilon_{37} \dot{r}$ equation, which can be written, using \eqref{3.2}, in the approximate form
\be\lab{3.3} \varepsilon_{37} \dot{r}=\frac{\de_{25}(\la_6s_3-\nu+\eps_{19}\nu g)}{\la_6}+\left(\frac{\de_{25}}{\la_6}-\de_3\right)rg-[r-1]_+.\ee
Figure \ref{fig4new} plots $\varepsilon_{37} \dot{r}$ as a function of $r$  for  the two values of $s_3$ used in figure \ref{fig3new}.
We see that for the normal value $s_3=0.782$, there is a stable steady state in which $r$ and thus $g$ are $O(1)$, and because of our choice of scales the deep ocean is anoxic. However, for $s_3<\dfrac{\nu}{\la_6}\approx 0.33$, $r$ collapses to zero, and the oxygen level increases dramatically. This sharp transition is due to the apparent parametric accident that $\dfrac{\de_{25}}{\la_6}-\de_3=0$, according to the values in \eqref{B2} and \eqref{B5}. It seems unlikely such a coincidence would occur, but in fact, working our way through the definitions of the parameters in the appendix,  we do find that
\begin{equation}\frac{\de_3\la_6}{\de_{25} }=1.\end{equation}
\begin{figure}
\begin{center}
\includegraphics[width=0.6\textwidth]{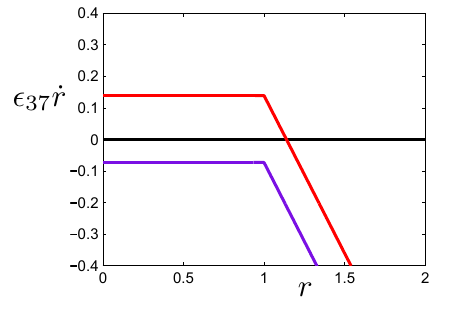}
\caption{$\varepsilon_{37} \dot{r}$ as a function of $r$ given by \eqref{3.3} using parameter values from the Slomp model and $\nu$ set to $1$ to represent a poorly-mixed ocean. In the upper curve, $s_3=0.782$ whereas in the lower curve, $s_3=0.1$. The quantities $\varepsilon_{37} \dot{r}$ and $r$ are dimensionless.}
\lab{fig4new}
\end{center}
\end{figure}
Ultimately, this is due to the equal coefficients $b_1$ in the rates of aerobic and anaerobic respiration in \eqref{2.1}. The model thus takes the very simple form in the anoxic case $\la_6s_3>\nu$:
\be \lab{anoxic_r_ode} \varepsilon_{37} \dot{r}=\de_{25} \left(s_3-\frac{\nu}{\la_6}+\frac{\eps_{19}\nu g}{\la_6}\right)-[r-1]_+.\ee
Anoxic equilibrium is obtained in a time scale $t\sim \eps_{37}\sim 10^{-2}$, corresponding to about 30 y.

\subsubsection{\lab{sec3.1}The oxic deep ocean}

What if $\la_6s_3<\nu$? It is then necessary to rescale the variables as
\be\lab{3.6}r\sim \eps_{19},\quad g\sim\frac{1}{\eps_{19}},\ee
and \eqref{3.1} now takes the approximate form (since $\eps_{19}\ll 1$)
\bea\lab{3.7}
 \frac{\varepsilon_{32}}{\eps_{19}} \dot{g} &=& \nu -\la_6s_3- \nu g  - r g,\dnon
 \varepsilon_{37}\eps_{19} \dot{r} &=&  -\delta_3 r g;
\eea
thus $r\to 0$ (approximately) very  rapidly, and then on a time scale of $t\sim\eps_{32}/\eps_{19}\sim 1$, corresponding to 3,000 y, $g\to 1-\frac{\la_6s_3}{\nu}$, and in dimensional terms, $0.33 \left(1-\dfrac{\la_6s_3}{\nu} \right)$ mM.

\subsubsection{\lab{sec3.2}Numerical verification}

\begin{figure}[!ht]
\centering
$\begin{array}{cc}
\includegraphics[width=0.45\textwidth]{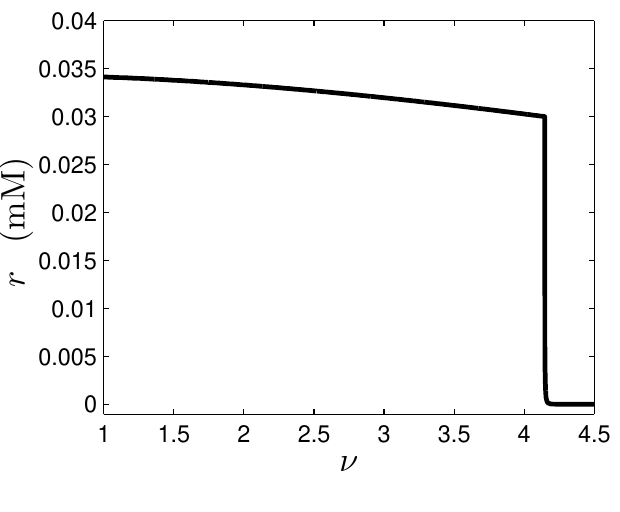} &

\includegraphics[width=0.45\textwidth]{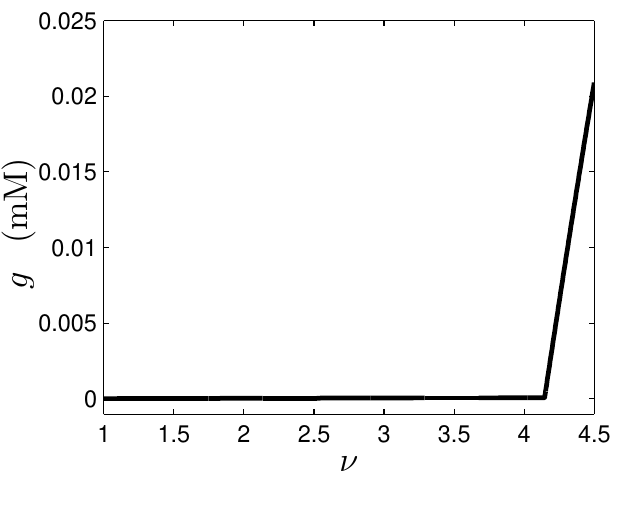}
\end{array}$

\caption{Steady-state values of $r$ and $g$ as a function of the dimensionless oceanic mixing parameter $\nu$. Solutions of \eqref{2.13} were obtained numerically before variables were converted to their dimensional forms.
 }
\label{varymix}
\end{figure}

We have provided a description of the dynamical behaviour of the oxygen subsystem in oxic and anoxic conditions as well as characterising the transition between the oxic and anoxic states. We will now assess the accuracy of these predictions through numerical solutions of \eqref{2.13}. Slomp and Van Cappellen (2007) showed that the mixing parameter could be varied to induce switches between oxic and anoxic conditions. Thus, in figure \ref{varymix}, we plot steady-state values of $g$ and $r$ as a function of the mixing parameter $\nu$. Note that here, and in the remainder of this section, we have reverted to dimensional variables for ease of interpretation.
Examining the numerical solutions, we note that at a critical value of $\nu \approx 4.14$, $r$ falls abruptly from $0.03$ mM to near zero while $g$ begins to increase rapidly.
Thus, the sudden shift in (equilibrium) redox state predicted once a critical parameter value has been exceeded (see section \ref{sec3}) appears to be borne out by the numerical solutions.

In view of the preceding discussion, one might expect to see a jump in $g$ at the same critical value of $\nu\approx 4.14$ in figure \ref{varymix}. This is masked by the fact that the jump in $g$ corresponding to the jump in $r$ is on the anoxic oxygen scale $\sim 3\times 10^{-5}$ mM, and thus not visible in figure \ref{varymix}. Further, we can see from (\ref{anoxic_r_ode}) that the jump in $r$ occurs when $\lambda_6s_3\approx \nu$, and that from (\ref{3.2}), the anoxic oxygen is $g\approx \dfrac{\lambda_{11}\nu}{\lambda_6s_3-\nu}$ when $r\approx 0$. So when $r$ jumps down, the anoxic-scaled $g$ increases rapidly, and this is indicated by the rapid rise in $g$ (on the oxic scale) in figure \ref{varymix}, which is proportional to $\nu-\lambda_6s_3$, as can be seen from (\ref{3.7}).

\begin{figure}[!ht]
\centering
\includegraphics[width=0.6\textwidth]{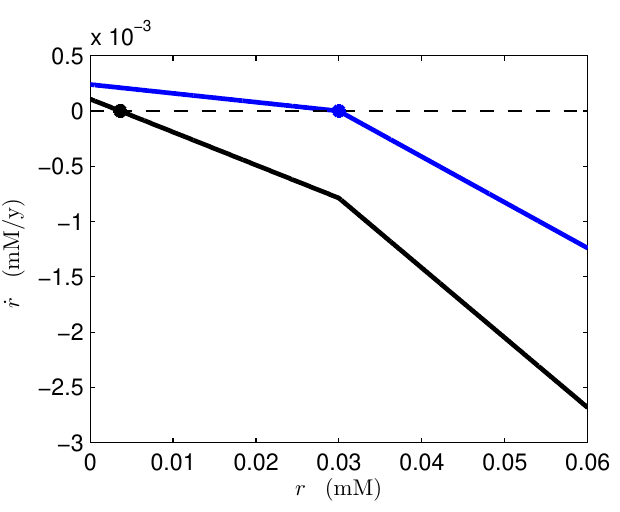}

\caption{$\dot{r}$ as a function of $r$ plotted at $\nu = 4.1457$ (blue curve) and $\nu = 4.1496$ (black curve). Plotted variables are in dimensional form. The marked points correspond to steady-state values of $r$ and all other system variables are at their numerically obtained steady-state values.}
\label{clines}
\end{figure}

To verify that we have successfully captured the mechanism behind this abrupt change, we use our numerical output to plot $\dot{r}$ as a function of $r$. We carry out this exercise on both sides of the apparent discontinuity with the results shown in figure \ref{clines}.
A small change in $\nu$ brings about a drastic shift in the position of the $\dot{r}$ curve and hence a large change in the equilibrium value of $r$. It is instructive to compare these curves with the dimensionless equivalents in figure \ref{fig4new} which have assumed $\nu=1$.
It appears that the mixing parameter is sufficiently high in figure \ref{clines} that the $\eps_{19} \nu g$ term in \eqref{3.3} is no longer negligible. This has the effect of converting the flat piece of $\dot{r}$ to a monotonically decreasing function of $r$.
Nonetheless, the relationship between $r$ and $\dot{r}$ at low $r$ values is relatively insensitive, facilitating the large shift in steady-state concentration as the mixing parameter moves through a critical threshold.

Finally, using $\nu=1$ and $\nu=4.5$ to represent anoxic and oxic oceans respectively, we examine the dynamics of the oxygen sub-system. We recall that, in section \ref{sec3}, we predicted that anoxic equilibrium would be obtained in a time scale of approximately 30 y. Meanwhile, in section \ref{sec3.1}, we predicted that a well-mixed deep ocean would recover its oxygen levels in a time scale of approximately 3000 y. In order to assess the validity of these estimates, we set all other variables to steady-state and plot numerical solutions for $g$ and $r$ with $\nu=1$ (see figure \ref{grtime}(i)) and $\nu=4.5$ (see figure \ref{grtime}(ii)). In both cases, we observe strong agreement between the numerical output and our analytical predictions.
    \begin{figure}
\centering

\includegraphics[width=\textwidth]{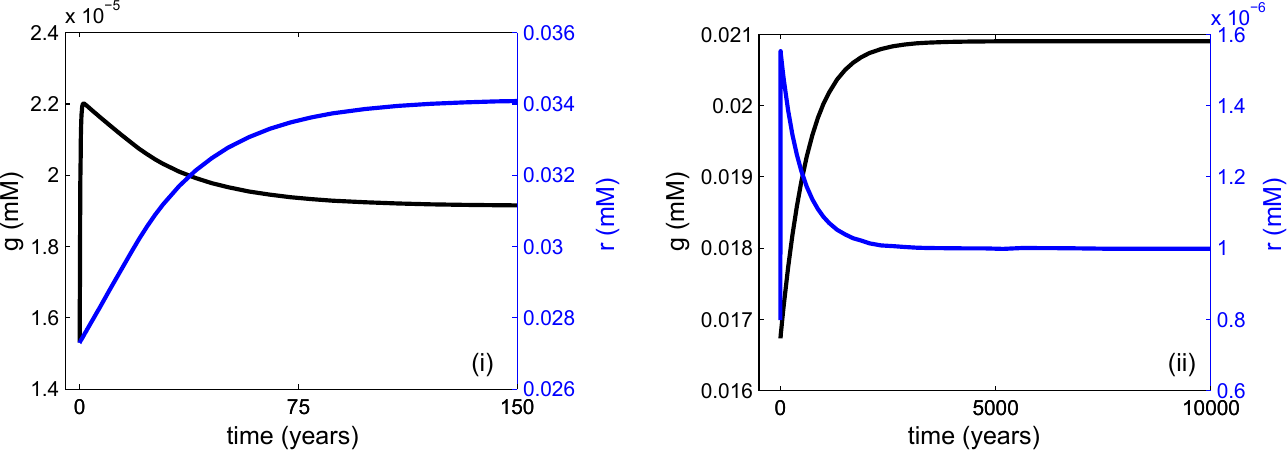}

\caption{ Equilibration of oxygen ($g$) and reduced substances ($r$) concentrations at two different mixing rates ($\nu$). In both cases, a steady state for the overall system of differential equations is first found numerically with the associated concentrations of oxygen and reduced substances then perturbed to 80\% of their steady-state values. In (i), $\nu=1$ and the ocean is poorly mixed whereas in (ii), $\nu=4.5$ and the ocean more closely resembles the present-day configuration. In both cases, we solve \eqref{2.13} numerically and then convert all variables to their dimensional forms. }

\lab{grtime}
\end{figure}

\section{\lab{sec4}Discussion of transition to anoxia}
\reseteqnos{4}

The analysis in section \ref{sec2.2} suggests that the chemical components in the different oceanic boxes rapidly (here meaning $\ll 3,\!000$ y) come to an approximate equilibrium, where the values are determined in terms of the deep ocean reactive phosphorus $s_4$, which however evolves over a much longer time scale $\sim 180,\!000$ y to an eventual equilibrium given by \eqref{2.20}. The surface ocean reactive phosphorus $s_3$ follows the same slow evolution, being determined by \eqref{2.19}. During this slow evolution of $s_3$, the deep ocean will rapidly (30 y) become anoxic if $\la_6s_3>\nu$, whereas if $\la_6s_3<\nu$ it becomes oxic, slightly less rapidly (3,000 y).

In section \ref{sec3}, we analysed the mechanisms responsible for shifts between anoxic and oxic deep oceans in the model. Starting with a poorly-mixed ocean, we observed that the processes of reoxidation of reduced substances and aerobic respiration appear in both the differential equation for oxygen and the differential equation for reduced substances. Equilibrium of the oxygen equation implies that the losses of oxygen to these two processes are effectively cancelled out by the net supply of oxygen from the surface ocean. This, in turn, means that the remaining two terms in the reduced-substances equation must balance at equilibrium. One of these terms is a constant (or weakly decreasing) input of reduced substances. The other term corresponds to the removal of reduced substances via precipitation (assumed to be followed by burial in sediment).
However, this precipitation, modelled by $k_{\text{prec}} \theta(r)$ in the dimensional system, is activated only when the concentration of reduced substances exceeds a prescribed threshold value. This non-smooth feature of the model produces a kink in the relationship between $\dot{r}$ and $r$ (see figure \ref{fig4new} and figure \ref{clines}). The presence of this kink means that large changes in equilibrium concentrations can be brought about by small changes in system parameters.

In simple biogeochemical terms, our analysis suggests that when surface ocean reactive phosphorus $s_3$ is too large, the deep ocean will become anoxic. Assuming the ocean is generally near steady-state conditions, we have a statement involving the equilibrium value of $s_3$. We compute this equilibrium value both numerically and using our analytical approximation (given by (\ref{2.19})) and plot $\la_6 s_3-\nu$ as a function of $\nu$ in figure \ref{thresholds}. We note that the upper-limit value of $\nu=10$ corresponds to the modern, well-mixed ocean. By comparison with the Slomp article's `degree of anoxicity' measure, we observe that this quantity successfully captures the deep ocean's transition from an oxic to an anoxic state at $\nu \approx 4$.
This model therefore has the capacity to explain ocean anoxia events, depending on the assumed parameters of the problem. It is thus important to unravel what all these complicated parameter combinations mean in terms of the dimensional parameters of the physical system.
While $\la_6$ is independent of the mixing parameter $\nu$, the value of $s_3$ depends on $\nu$ as well as other system parameters. This functional dependence is not known exactly.
However, by using the approximate form of the $s_3$ steady-state value, we can express $\la_6 s_3-\nu$ as an explicit function of the model's dimensional parameter set.

    \begin{figure}
\centering

\includegraphics[width=0.6\textwidth]{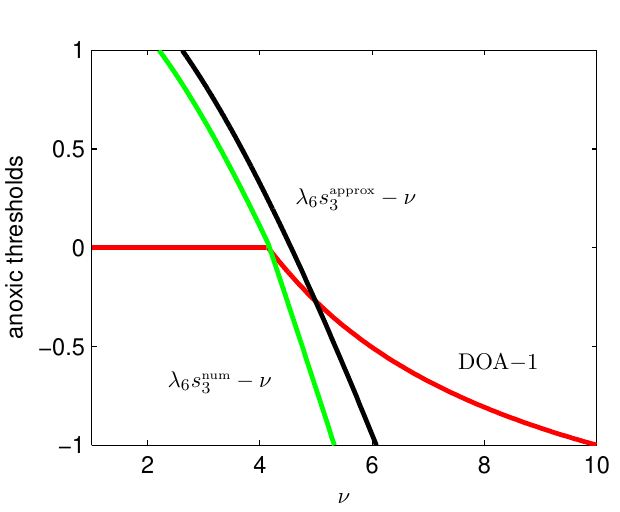}

\caption{ Anoxia-onset metrics as a function of the mixing parameter $\nu$. The quantity $\la_6 s_3-\nu$ is presented in two forms, one using a numerical estimate of the $s_3$ steady-state value ($\la_6 s_3^{\mbox{\tiny num}}-\nu$) and one using an analytical approximation ($\la_6 s_3^{\mbox{\tiny approx}}-\nu$). Negative values of the metrics correspond to oxic conditions.
The degree of anoxicity (DOA) quantity, defined as $1 - \frac{g}{g_0^*}$ in the original Slomp paper, is shown as a reference. Note that oxygen levels, though extremely small at low values of the mixing parameter, remain nonzero and therefore, the DOA never equals one. }
\lab{thresholds}
\end{figure}

Unfortunately, although the analysis is simple, the dependence of the critical parameter on the physical inputs is non-trivial in the extreme, to the extent that in the appendix we give an algorithm to compute $\la_6 s_3^{\mbox{\tiny approx}}-\nu$, and in the electronic supplementary material provide a code which does this (see Online Resource 1).
The fully expanded expression for $\la_6 s_3^{\mbox{\tiny approx}}-\nu$ depends on $48$ dimensional parameters. Here, we focus on the influence of $a_{53}$, the riverine input of SRP. Slomp and Van Cappellen's (2007) numerical exploration revealed that anoxia would occur if the present ocean's circulation rate was reduced by 50\% ($\nu=5$ in our notation) while the supply of reactive phosphorus from the continents was simultaneously boosted by 20\%.
They suggested that such an increase could be caused by coastal erosion linked to sea level rise.

Setting $\nu=5$ and leaving all other parameters at their previously assigned values, we plot $\la_6 s_3^{\mbox{\tiny approx}}-\nu$ as a function of $a_{53}$. Figure \ref{b53fig} demonstrates that $\la_6 s_3^{\mbox{\tiny approx}}-\nu$ switches from negative to positive as $a_{53}$ is increased with the crossover occurring when $a_{53}$ is 6\% higher than its baseline value.
Numerical study (not shown) reveals that the threshold actually lies at a value of $a_{53}$ that is 12\% higher than the baseline value (i.e., between our prediction and the value used by Slomp and Van Cappellen).
Thus, the quantity $\la_6 s_3^{\mbox{\tiny approx}}-\nu$ appears to be able to predict changes in ocean oxygen status, whether they be linked to circulation or other factors.

    \begin{figure}
\centering
\includegraphics[width=0.6\textwidth]{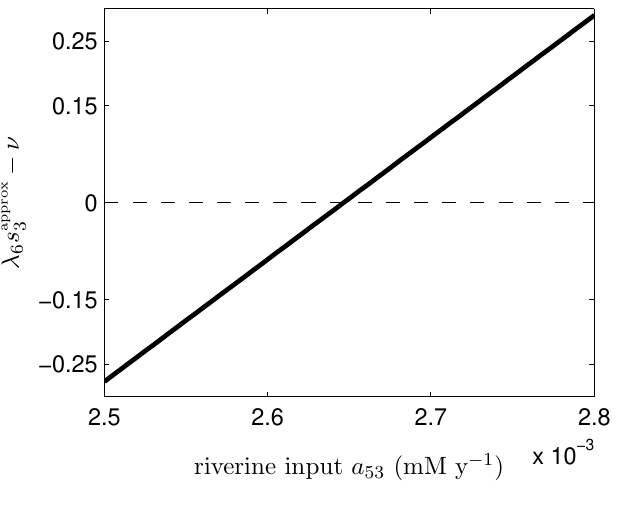}
\caption{ Anoxia onset metric as a function of the riverine input parameter $a_{53}$ where we have set $\nu=5$. Positive values of $\la_6s_3-\nu$ correspond to anoxic conditions and negative values to oxic. }
\lab{b53fig}
\end{figure}

It is important to emphasise that we focused on changes in $\nu$ and $a_{53}$ only because Slomp and Van Cappellen's (2007) numerical exploration examined these two factors. The purpose of producing an algebraic expression for $\la_6 s_3^{\mbox{\tiny approx}}-\nu$ was to understand how the transition to anoxia depends on system parameters, more broadly.
This `All-At-A-Time' approach to sensitivity analysis (Pianosi \etal 2016) allows us to assess the robustness of the model output to the modelling assumptions while avoiding the need to produce a high volume of model runs and then visually compare model predictions.

\section{\lab{sec5}Conclusions}
\reseteqnos{5}

In this article, we have systematically analysed the model of the phosphorus cycle in the ocean given by Slomp and Van Cappellen (2007).
Through careful scaling of the Slomp model, we identified a large number of negligible steady-state fluxes. We also isolated distinct time scales associated with system equilibration. By exploiting these two factors, we were able to effectively decouple the subsystem of oxygen and reduced substances from the carbon-phosphorus cycle.

While soluble reactive phosphorus acts as an (effectively static) input to the oxygen subsystem, the contribution of oxygen to the cycling of carbon and phosphorus can be safely ignored. In particular, this means that a range of nonlinear, non-smooth functions used to model redox dependence in the burial of sorbed P, particulate organic P and particulate organic carbon can be excised without affecting our qualitative findings. From a starting point of eighteen nonlinear equations, we separately analysed a set of sixteen (approximately) linear equations which govern carbon and phosphorus dynamics and a pair of equations which explain the chemistry of the oxic deep ocean, the chemistry of the anoxic deep ocean and the nature of the transition between the two.

Having partitioned the system into two parts, we can elucidate the nature of the transition between oxic and anoxic oceans. A small change in system parameters produces abrupt, almost discontinuous, switches in the equilibrium concentrations of oxygen and reduced substances. We link this sensitivity in the model to the functional form prescribed for the removal of reduced substances as solid phases and the functional form for microbial respiration. Allison and Martiny (2008) refer to this kind of microbial model as a ``black box'' with ``microorganisms buried within equation structure as kinetic constants and response functions''. Our analysis highlights the need to compare the predictions of such studies with those of models that explicitly incorporate microbial biomass, in order to enhance understanding of how anoxia occurs.

With the nature of transition to anoxia established, we sought to determine the system parameters responsible for driving such a transition.
Unfortunately, due to the scope of the original model (containing $69$ parameters), the critical controlling parameter defies succinct characterisation.
However, by focusing on a small subset of the system parameters (i.e., mixing rate and riverine input), we demonstrated that one can accurately predict the outcome of changes in the the rate of a given process. While our focus here was on providing this kind of proof-of-concept, future work could entail analytic study of the expression \eqref{lam6s3expr} in the appendix. In particular, one can explicitly determine whether the ocean's oxygen status is affected by variation (or covariation) in a few parameters of interest. More generally, we suggest that this article demonstrates the viability of adopting a systematic, mathematical approach in studying the behaviour of large biogeochemical models. The deduction of parameterised steady-state concentrations and equilibration time scales, as we have presented here, is generally beyond the reach of a purely computational approach to biogeochemistry.

\subsection*{Acknowledgements}

We acknowledge the support of the Mathematics Applications
Consortium for Science and Industry ({\tt www.macsi.ul.ie}) funded by the
Science Foundation Ireland mathematics  grant 12/1A/1683. %
This publication has emanated from research conducted with the financial support of Science Foundation Ireland under Grant Number: SFI/09/IN.1/I2645.
No new data was used in this study.

\resetsectnos{A}
\section*{Appendix}
\reseteqnos{A}

\subsection*{Non-dimensionalisation}

We non-dimensionalise the model as described before \eqref{2.7}. In the model as presented by Slomp and Van Capellen (2007) the chemical quantities are measured in Tmol and the reservoir volumes in Tm\up{3}. We have written the model equations \eqref{2.1} in terms of concentrations, which thus have the units mol m$^{-3} =$ mM (millimolar). The  set of scales is given by (in units of mM)
\bea\lab{B4}
 {[r]} = C_{RP} \approx 3 \times 10^{-2}, &\qquad& [s_1] = \frac{a_{53}}{d_{16}} \approx 2.59 \times 10^{-4},\dnon
  {[c_1]} = \frac{b_3}{b_4} [s_1] \approx 1.21 \times 10^{-1},&\qquad&
 {[c_2]} = \frac{b_6}{b_7} [s_2] \approx 4.30 \times 10^{-3}, \dnon
   {[s_3]} = d_{18} [s_2] \approx 5.98 \times 10^{-7},&\qquad&{[s_2]} = d_{17} [s_1] \approx 8.51 \times 10^{-5},\dnon
 {[c_3]} = \frac{b_8}{b_{10}} [s_3] \approx 4.79 \times 10^{-5},&\qquad&
 {[c_4]} = \frac{b_{12}}{b_{13}} [s_3] \approx 2.72 \times 10^{-5}, \dnon
 {[p_1]} = \frac{m_{72}}{a_{18}} [s_1] \approx 1.14 \times 10^{-3},&\qquad& [f_1] = \frac{m_{73}}{a_{20}} [s_1] \approx 2.00 \times 10^{-4},\dnon
   {[p_2]} = \frac{m_{75}}{a_{27}} [s_2] \approx 4.06 \times 10^{-5},&\qquad&
 {[f_2]} = \frac{m_{76}}{a_{29}} [s_2] \approx 1.80 \times 10^{-6},\dnon
 {[p_3]} = \frac{m_{77}}{a_{36}} [s_3] \approx 4.52 \times 10^{-7},&\qquad& [f_3] = \frac{m_{78}}{a_{38}} [s_3] \approx 8.60 \times 10^{-9},\dnon
 {[s_4]} = \frac{a_{39}}{  m_{58} v_a  } [p_4] \approx 7.03 \times 10^{-6} ,&\qquad& [p_4] = \frac{b_{42}}{R_{CP} a_{43}} [s_3] \approx 2.56 \times 10^{-7},\dnon
 {[f_4]} = \frac{a_{44}}{a_{45}} [f_3] \approx 3.30 \times 10^{-10},&\qquad& {[g]} = \frac{10^{6} m_{0} v_a g_s}{k_{\text{redox}}[r]} \approx 3.47 \times 10^{-5}.
\eea
where
\be\lab{A2} d_{16} = m_{71}-\frac{ a_{14} m_{72} }{ a_{18} },\quad
d_{17} = \frac{ a_{21} a_{27} }{ m_{74} a_{27}-a_{22} m_{75} } , \quad d_{18} = \frac{ a_{31} a_{36} }{ b_{32} a_{36}-a_{30} m_{77} }.\ee
The dimensionless parameters of the resulting model \eqref{2.7} are given by the following three sets. First the $O(1)$ parameters $\la_i$:
\begin{eqnarray}\lab{B1}
 \lambda_1 = \frac{a_{53}}{m_{71} [s_1]} \approx 0.238 , && \lambda_2 = \frac{a_{14} [p_1]}{m_{71} [s_1]} \approx 0.762, \quad
 \lambda_{3}= \frac{a_{22} [p_2]}{ m_{74} [s_2] } \approx 0.971,\dnon
\lambda_{4}= \frac{a_{30} [p_3]}{ b_{32} [s_3] } \approx 0.853,&&
 \lambda_{5} = \frac{a_{59}}{ [s_4] m_{58} v_a } \approx 2.31, \quad  \lambda_{11} = \frac{K_m}{[g]} \approx 2.88, \dnon
 \lambda_{20}= \frac{a_{82}  v_a [s_2]}{ b_{32} [s_3] } \approx 0.15. &&  \quad
 \end{eqnarray}
 Next, the small, but not very small, parameters $\de_i$:
 \begin{eqnarray}\lab{B2}
\delta_{1}= \frac{m_{56}  v_a [s_4]}{ b_{32} [s_3] } \approx 0.124, && \delta_{2}= \frac{a_{31} [s_2]}{ b_{32} [s_3] } \approx 0.147,\quad
 \delta_{3}= \frac{ k_{\text{redox}} [r_4] [g_4]}{ 10^6 k_{\text{prec}} } \approx 0.104, \dnon
\delta_{4}= \frac{b_{5} [c_1]}{ b_{7} [c_2]  } \approx 0.13 && \delta_{5}= \frac{a_{26} [p_1]}{ a_{27} [p_2]  } \approx 0.132. \quad
 \end{eqnarray}
\no Finally, the very small parameters $\eps_i$:
 \begin{eqnarray}\lab{B3}
\varepsilon_{1}= \frac{a_{15} [f_1]}{ [s_1] m_{71} } \approx 9.52 \times 10^{-3}, &\qquad&
\varepsilon_{2}= \frac{a_{21} [s_1]}{ m_{74} [s_2] } \approx 2.92 \times 10^{-2} ,\dnon
\varepsilon_{3}= \frac{a_{23} [f_2]}{ m_{74} [s_2] } \approx 9.89 \times 10^{-3}, &\qquad&
\varepsilon_{4}= \frac{m_{54}  v_a [s_4]}{ m_{74} [s_2] } \approx 8.11 \times 10^{-4},\dnon
\varepsilon_{6}= \frac{a_{40} [f_4]}{ [s_4] m_{58}  v_a } \approx 7.33 \times 10^{-2}, &\qquad&
\varepsilon_{8}= \frac{a_{34} [p_2]}{ [p_3] a_{36} } \approx 8.22 \times 10^{-2},\dnon
\varepsilon_{9}= \frac{b_{35} [c_2]}{ [p_3] a_{36} } \approx 1.12 \times 10^{-2},  &\qquad&
\varepsilon_{10}= \frac{b_{85} b_{41} [c_2]}{ (C/P)_{\mbox{\tiny oxic}} [p_4] a_{43} } \approx 1.01 \times 10^{-4}, \dnon
\varepsilon_{11}= \frac{b_{85} b_{42} [s_3]}{ (C/P)_{\mbox{\tiny oxic}} [p_4] a_{43} } \approx 1.44 \times 10^{-3},  &\qquad&
\varepsilon_{13}= \frac{b_{9} [c_2]}{ [c_3] b_{10} } \approx 7.01 \times 10^{-2}, \dnon
\varepsilon_{14}= \frac{b_{11} [c_2]}{ [c_4] b_{13} } \approx 7.01 \times 10^{-2},&\qquad&
\varepsilon_{15}= \frac{b_{11} [c_2] b_{85}}{ [c_4] b_{13} } \approx 2.26 \times 10^{-4}, \dnon
\varepsilon_{16}= \frac{b_{12} b_{85} [s_3]}{ [c_4] b_{13} } \approx 3.22 \times 10^{-3}, &\qquad&
     \varepsilon_{19}= \frac{[g_4]}{ g_s } \approx 1.07 \times 10^{-4},\dnon
 \varepsilon_{20} = \frac{m_{58} v_a }{m_{71}} \approx 7.89 \times 10^{-6},&\qquad&
  \varepsilon_{21} = \frac{m_{58} v_a }{m_{74}} \approx 2.99 \times 10^{-4}, \dnon
 \varepsilon_{22} = \frac{m_{58} v_a }{b_{32}} \approx 4.46 \times 10^{-4},&\qquad&
 \varepsilon_{23} = \frac{m_{58} v_a }{a_{18}} \approx 3.79 \times 10^{-5}, \dnon
 \varepsilon_{24} = \frac{m_{58} v_a  }{a_{27}} \approx 1.46 \times 10^{-4},&\qquad&
  \varepsilon_{25} = \frac{m_{58} v_a }{a_{36}} \approx 3.95 \times 10^{-4}, \dnon
\varepsilon_{26} = \frac{m_{58} v_a }{a_{43}} \approx 3.63 \times 10^{-2}, &\qquad&
 \varepsilon_{27} = \frac{m_{58} v_a }{a_{20}} \approx 6.41 \times 10^{-4} ,\dnon
 \varepsilon_{28} = \frac{m_{58} v_a  }{b_{4}} \approx 4.02 \times 10^{-5} ,&\qquad&
 \varepsilon_{29} = \frac{m_{58} v_a }{b_{7}} \approx 1.45 \times 10^{-4},\dnon
 \varepsilon_{30} = \frac{m_{58} v_a }{b_{10}} \approx 3.90 \times 10^{-4},&\qquad&
\varepsilon_{31} = \frac{m_{58} v_a }{b_{13}} \approx 3.64 \times 10^{-2} ,\dnon
\varepsilon_{32} = \frac{m_{58} [g] v_a }{m_{0} g_s} \approx 1.07 \times 10^{-4},&\qquad&
 \varepsilon_{33}= \frac{b_{41} [c_2]}{ [p_4] a_{43} R_{CP} } \approx 7.01 \times 10^{-2}, \dnon
 \varepsilon_{34} = \frac{m_{58} v_a }{a_{29}} \approx 6.41 \times 10^{-4}&\qquad&
 \varepsilon_{35} = \frac{m_{58} v_a }{a_{38}} \approx 6.41 \times 10^{-4} \dnon
 \varepsilon_{36} = \frac{m_{58} v_a }{a_{45}} \approx 6.41 \times 10^{-4}&\qquad&
 \varepsilon_{37} = \frac{m_{58} v_a [r]}{k_{\text{prec}}} \approx 9.61 \times 10^{-3} \dnon
 \varepsilon_{38}= \frac{ b_{1} [c_4]}{ k_{\text{prec}} } \approx 3.12 \times 10^{-4},&\qquad&
 \varepsilon_{39}= \frac{b_{1} [c_4]}{ m_{0} v_a g_s } \approx 2.99 \times 10^{-3},\dnon
\varepsilon_{99}= \frac{b_{86} v_a [c_2]}{ [p_4] a_{43} R_{CP} } \approx 7.2  \times 10^{-2}, &\qquad&
\varepsilon_{101}= \frac{a_{74} v_a }{ m_{74} } \approx 9.81 \times 10^{-3},\dnon
\varepsilon_{102}= \frac{a_{81} v_a  }{ a_{27} } \approx 4.79 \times 10^{-3}, &\qquad&
\varepsilon_{103}= \frac{b_{81}  v_a }{ b_{7} } \approx 4.75 \times 10^{-3},\dnon
\varepsilon_{104}= \frac{b_{85} b_{86} [c_2] v_a }{ (C/P)_{\mbox{\tiny oxic}} [p_4] a_{43} } \approx 1.03 \times 10^{-4}, &\qquad&
\varepsilon_{105}= \frac{b_{83} [c_2] b_{85} v_a  }{ [c_4] b_{13} } \approx 2.31 \times 10^{-4}, \dnon
\end{eqnarray}

\begin{eqnarray*}
\varepsilon_{106}= \frac{m_{32} v_a }{ b_{32} } \approx 1.2 \times 10^{-2}, &\qquad&
\varepsilon_{107}= \frac{[s_3]}{ [s_4] } \approx 8.5 \times 10^{-2}, \dnon
\varepsilon_{110}= \frac{a_{83} v_a [p_2]}{ [p_3] a_{36} } \approx 8.4 \times 10^{-2}, &\qquad&
\varepsilon_{111}= \frac{b_{84} v_a [c_2]}{ [p_3] a_{36} } \approx 1.2 \times 10^{-2}, \dnon
\varepsilon_{112}= \frac{b_{82} v_a [c_2]}{ [c_3] b_{10} } \approx 7.2 \times 10^{-2},  &\qquad&
\varepsilon_{113}= \frac{b_{83} v_a [c_2]}{ [c_4] b_{13} } \approx 7.2 \times 10^{-2}. \dnon
\end{eqnarray*}

When the rescaling introduced before \eqref{2.13} is done, the  new dimensionless  parameters are given by
\begin{eqnarray}\lab{B5}
 \lambda_6 = \varepsilon_{39} [\bar{s}_3] \approx 2.988, &\qquad& \delta_{25} = \varepsilon_{38} [\bar{s}_3] \approx 0.311,\dnon
 \varepsilon_{40} = \frac{\delta_4}{ [\bar{s}_2]} \approx  3.8 \times 10^{-3}, &\qquad& \varepsilon_{41}= \varepsilon_{13} \frac{[\bar{s}_2]}{[\bar{s}_3]} \approx 2.4 \times 10^{-3},\dnon
 \varepsilon_{42} = \varepsilon_{14} \frac{[\bar{s}_2]}{[\bar{s}_3]} \approx 2.4 \times 10^{-3}, &\qquad& \varepsilon_{43} = \varepsilon_{15} \frac{[\bar{s}_2]}{[\bar{s}_3]} \approx 7.77 \times 10^{-6},\dnon
 \varepsilon_{44} = \frac{\varepsilon_{2}}{[\bar{s}_2]} \approx  8.51 \times 10^{-4}, &\qquad& \varepsilon_{45} = \varepsilon_{4} \frac{[\bar{s}_3]}{[\bar{s}_2]} \approx 2.36 \times 10^{-2}, \dnon
 \varepsilon_{46} = \frac{\delta_5}{[\bar{s}_2]} \approx 3.8 \times 10^{-3}, &\qquad& \varepsilon_{47} = \delta_{2} \frac{[\bar{s}_2]}{[\bar{s}_3]} \approx 5.1 \times 10^{-3},\dnon
 \varepsilon_{48} = \varepsilon_{8} \frac{[\bar{s}_2]}{[\bar{s}_3]} \approx 2.8 \times 10^{-3}, &\qquad& \varepsilon_{49} = \varepsilon_{9} \frac{[\bar{s}_2]}{[\bar{s}_3]} \approx 3.87 \times 10^{-4},\dnon
  \varepsilon_{50} = \frac{\lambda_{5}}{[\bar{s}_3]} \approx 2.3 \times 10^{-3}, &\qquad& \varepsilon_{51} = \varepsilon_{33} \frac{[\bar{s}_2]}{[\bar{s}_3]} \approx 2.4 \times 10^{-3},\dnon
  \varepsilon_{52} = \varepsilon_{10} \frac{[\bar{s}_2]}{[\bar{s}_3]} \approx 3.48 \times 10^{-6}, &\qquad& \varepsilon_{120}=\frac{\varepsilon_{99} [\bar{s}_2]}{[\bar{s}_3]} \approx 2.5 \times 10^{-3},\dnon
        \varepsilon_{121}=\frac{\varepsilon_{110} [\bar{s}_2]}{[\bar{s}_3]} \approx 2.9 \times 10^{-3}, &\qquad&
        \varepsilon_{122}=\frac{\varepsilon_{111} [\bar{s}_2]}{[\bar{s}_3]} \approx 3.95 \times 10^{-4},\dnon
\varepsilon_{123}= \frac{\varepsilon_{112} [\bar{s}_2]}{[\bar{s}_3]} \approx 2.5 \times 10^{-3}, &\qquad&
        \varepsilon_{124}=\frac{\varepsilon_{113} [\bar{s}_2]}{[\bar{s}_3]} \approx 2.5 \times 10^{-3},\dnon
        \varepsilon_{125}=\frac{\varepsilon_{104} [\bar{s}_2]}{[\bar{s}_3]} \approx 3.55 \times 10^{-6}, &\qquad&
        \varepsilon_{126}=\frac{\varepsilon_{105} [\bar{s}_2]}{[\bar{s}_3]} \approx 7.94 \times 10^{-6},\dnon
        \varepsilon_{127}=\frac{\lambda_{20} [\bar{s}_2]}{[\bar{s}_3]} \approx 5.2 \times 10^{-3}. &&
 \end{eqnarray}

\subsection*{Dimensional parameters}

In this section, we present the definitions and values of all dimensional constants associated with the model \eqref{2.1}.
Before we do so, we must first document the parameters of the original Slomp model. The values assigned to the parameters of the Slomp model, as well as their units and physical interpretations, are listed in tables \ref{table4} and \ref{table5}. These parameters combine to produce the dimensional constants used in \eqref{B4} and the model \eqref{2.1}. The definitions of these dimensional constants are given in tables \ref{table3part1} and \ref{table3part2}.

\begin{table}[!t]
\begin{center}
\centering
\begin{tabular}{|l|l|l|l|}
\hline
Parameter & Value & Units & Description \\
\hline

$W_1$  & $3.6 \times 10^{13}$ & m$^3$ & Proximal coastal reservoir size \\
$W_2$  & $3.6\times 10^{15}$ & m$^3$ & Distal coastal reservoir size \\
$W_3$  & $4.98\times 10^{16}$ & m$^3$ & Surface ocean reservoir size \\
$W_4$  & $1.30\times 10^{18}$ & m$^3$ & Deep ocean reservoir size \\
$Wk_1$  & $3.70\times 10^{13}$ & m$^3/y$ & River input flux \\
$Wk_5$  & $3.78\times 10^{15}$ & m$^3/y$ & Ocean upwelling flux \\
$Wk_6$  & $3.78\times 10^{14}$ & m$^3/y$ & Coastal upwelling flux \\
$v_o$  & $v$ & & Mixing parameter \\
$v_c$  & $v$  & & Mixing parameter \\

$Ck_1$ &   $3.87 \times 10^1$ & $y^{-1}$   & Primary production rate in $W_1$ \\
$Ck_2$ &   $6.94$ & $y^{-1}$  & POC mineralisation rate in $W_1$ \\
$Ck_3$ &  $9.06\times 10^{-2}$ &  & POC burial parameter in $W_1$ \\
$Ck_4$ &  $1$ &  & POC export parameter ($W_1$ to $W_2$) \\
$Ck_5$ &  $1.06$ & $y^{-1}$  & Primary production rate in $W_2$ \\
$Ck_6$ &  $2.20$ & $y^{-1}$  & POC mineralisation rate in $W_2$ \\
$Ck_7$ &  $2.7/(560.25+4.66)$ &  & POC burial parameter in $W_2$ \\
$Ck_8$ &  $1$ &  & POC export parameter ($W_2$ to $W_3$) \\
$Ck_9$ &  $07.19\times 10^{-1}$ & $y^{-1}$  & Primary production rate in $W_3$ \\
$Ck_{10}$ &  $8.21\times 10^{-1}$ & $y^{-1}$  & POC mineralisation rate in $W_3$ \\
$Ck_{11}$ &  $(496.6/(3600+28.0125))$ &  & POC export parameter ($W_3$ to $W_4$)  \\
$Ck_{12}$ &  $8.81\times 10^{-3}$ & $y^{-1}$   & POC respiration rate in $W_4$ \\

$R_{CP}$ &  $1.06\times 10^{2}$ &  & Redfield ratio of Carbon and Phosphorus \\
$R_{CO_2}$ & $106/138$ & & Redfield ratio of Carbon and Oxygen \\
$(C/P)_{\mbox{\tiny oxic}}$ &  $237$ &  & C/P ratio of SOM (oxic conditions) \\
$(C/P)_{\mbox{\tiny anoxic}}$ &  $1100$ &  & C/P ratio of SOM (anoxic conditions) \\

$[{\rm RS}]_0$ & 0.03 & & Threshold RS concentration for precipitation \\
$Pk_1$ &  $9\times 10^{10}$ & mol/y$^{1}$   & SRP river input flux to $W_1$ \\
$Pk_2$ &  $9.25\times 10^{-1}$ & y$^{-1}$  & FeP burial rate in $W_1$ \\
$Pk_{3}$ &  $5.66\times 10^{-2}$ &  & CaP burial parameter in $W_1$ \\
$Pk_4$ &  $1$ &  & SRP export parameter ($W_1$ to $W_2$) \\
$Pk_{5a}$ &  $1\times 10^{-2}$ &  & Fish production parameter in $W_1$ \\
$Pk_{5b}$ &  $5\times 10^{-1}$ & y$^{-1}$  & Fish dissolution rate in $W_1$ \\
$Pk_{5c}$ &  $0$ & mol/y$^{1}$  & Fish burial flux in $W_1$ \\
$Pk_{6}$ &  $7.43$ & y$^{-1}$  & POP mineralisation rate in $W_1$ \\
$Pk_7$ &  $1$ &  & POP burial parameter in $W_1$ \\
\hline
\end{tabular}
\end{center}
\caption{Definition of the parameters of the Slomp model (continued below) }
\lab{table4}
\end{table}

\begin{table}[!t]
\begin{center}
\centering
\begin{tabular}{|l|l|l|l|}
\hline
Parameter & Value & Units & Description \\
\hline
$Pk_8$ &  $1$ &  & POP export parameter ($W_1$ to $W_2$) \\
$Pk_9$ &  $1.35\times 10^{-3}$ & y$^{-1}$  & FeP burial rate in $W_2$ \\
$Pk_{10}$ &  $2.70\times 10^{-3}$ &  & CaP burial parameter in $W_2$ \\
$Pk_{11}$ &  $1$ &  & SRP export parameter ($W_2$ to $W_3$) \\
$Pk_{12a}$ &  $1\times 10^{-2}$ &  & Fish production parameter in $W_2$ \\
$Pk_{12b}$ &  $5\times 10^{-1}$ & y$^{-1}$  & Fish dissolution rate in $W_2$ \\
$Pk_{12c}$ &  $0$ & mol/y$^{1}$  & Fish burial flux in $W_2$ \\
$Pk_{13}$ &  $2.18$ & y$^{-1}$   & POP mineralisation rate in $W_2$ \\
$Pk_{14}$ &  $0.00675/(0.044+5.28538)$ &  & POP burial parameter in $W_2$ \\
$Pk_{15}$ &  $1$ &  & POP export parameter ($W_2$ to $W_3$) \\
$Pk_{19}$ &  $8.11\times 10^{-1}$ & y$^{-1}$  & POP mineralisation rate in $W_3$ \\
$Pk_{20}$ &  $1\times 10^{-2}$ &  & Fish production parameter in $W_3$ \\
$Pk_{21}$ &  $0$ & mol/y  & Fish dissolution flux in $W_3$ \\
$Pk_{23}$ &  $5\times 10^{-1}$ & y$^{-1}$   & Fish sinking rate from $W_3$ to $W_4$ \\
$Pk_{24}$ &  $8.83\times 10^{-3}$ & y$^{-1}$  & POP mineralisation rate in $W_4$ \\
$Pk_{25}$ &  $5\times 10^{-1}$ & y$^{-1}$  & Fish dissolution rate in $W_4$ \\
$Pk_{26}$ &  $6.75\times 10^9$ & mol/y  & Maximum/oxic FeP burial flux in $W_4$ \\
$Pk_{27}$ &  $2.89\times 10^{-3}$ &  & CaP burial parameter in $W_4$ \\
$Pk_{28}$ &  $1.6/496.6$ &  & POP burial parameter in $W_4$ \\
$Pk_{29}$ &  $0$ &  & Fish burial parameter in $W_4$ \\

\hline
\end{tabular}
\end{center}
\caption{Definition of the parameters of the Slomp model, continued.}
\lab{table5}
\end{table}

\clearpage

\begin{table}[!ht]
\begin{center}
\centering
\begin{tabular}{|l|l|l|l|}
\hline
Constant & Value & Units & Definition \\
\hline
$a_{14}$ & $7.01  $ & y$^{-1}$ & $(1-{Pk}_3){Pk}_6 $\\
$a_{15}$ & $5 \times 10^{-1}$ & y$^{-1}$ & ${Pk}_{5b} $\\
$a_{18}$ & $8.46  $ & y$^{-1}$ &  $ ( {Pk}_6 W_1 + {Pk}_8{Wk_1})/W_1$\\
$a_{20}$ & $5 \times 10^{-1}$ & y$^{-1}$ &  $ {Pk}_{5b}$\\
$a_{21}$ & $1.03 \times 10^{-2}$ & y$^{-1}$ & $({Pk}_4{Wk_1})/W_2 $ \\
$a_{22}$ & $2.18  $ & y$^{-1}$ &  $(1-{Pk}_{10}){Pk}_{13} $\\
$a_{23}$ & $5 \times 10^{-1}$ & y$^{-1}$ &  ${Pk}_{12b} $\\
$a_{26}$ & $1.03 \times 10^{-2}$ & y$^{-1}$ & ${Pk}_8{Wk_1}(1 - {Pk}_{14})/W_2 $\\
$a_{27}$ & $2.19$ & y$^{-1}$ & $Pk_{13}+Pk_{15} Wk_1 / W_2$\\
$a_{29}$ & $5 \times 10^{-1}$ & y$^{-1}$ & ${Pk}_{12b}$   \\
$a_{30}$ & $8.11 \times 10^{-1}$ & y$^{-1}$ & ${Pk}_{19} $ \\
$a_{31}$ & $7.43 \times 10^{-4}$ & y$^{-1}$ & $Pk_{11} Wk_1 / W_3$ \\
$a_{34}$ & $7.43 \times 10^{-4}$ &y$^{-1}$ & $Pk_{15} Wk_1 / W_3$ \\
$a_{36}$ & $8.11 \times 10^{-1}$ & y$^{-1}$& ${Pk}_{19}$\\
$a_{38}$ & $5 \times 10^{-1}$ & y$^{-1}$ & ${Pk}_{23}$ \\
$a_{39}$ & $8.80 \times 10^{-3}$ &y$^{-1}$ & $ {Pk}_{24}(1 - {Pk}_{27})$\\
$a_{40}$ & $5 \times 10^{-1}$ &y$^{-1}$  & $ {Pk}_{25}$\\
$a_{43}$ & $8.23 \times 10^{-3}$ &y$^{-1}$  & ${Pk}_{24} $ \\
$a_{44}$ & $1.92 \times 10^{-2}$ &y$^{-1}$ & $\big( {Pk}_{23}(1-{Pk}_{29})W_3\big)/W_4$ \\
$a_{45}$ & $5 \times 10^{-1}$ &y$^{-1}$  & $ {Pk}_{25}$ \\
$a_{53}$ & $2.5 \times 10^{-3}$ & mM y$^{-1}$ & ${Pk}_1/W_1 $\\
$a_{59}$ & $5.20 \times 10^{-9}$ & mM y$^{-1}$& $ {Pk}_{26}/W_4 $\\
$a_{74}$ & $1.05 \times 10^{-1}$ & y$^{-1}$ & $Pk_{11} Wk_6 / W_2$ \\
$a_{81}$ & $1.05 \times 10^{-1}$ & y$^{-1}$ & $Pk_{15} Wk_6 / W_2$ \\
$a_{82}$ & $7.56 \times 10^{-3}$ & y$^{-1}$ & $Pk_{11} Wk_6 / W_3$ \\
$a_{83}$ & $7.56 \times 10^{-3}$ &y$^{-1}$ & $Pk_{15} Wk_6 / W_3$ \\
$b_1$ & $1.15 \times 10^{-2}$ & y$^{-1}$ & ${Ck}_{12}/R_{CO_2}$\\
$b_3$ & $3.73 \times 10^{3}$ & y$^{-1}$ & $(1-{Ck}_3)\times {Ck}_1 \times R_{CP}$\\
$b_4$ & $7.97  $ & y$^{-1}$ & $({Ck}_2 W_1 + {Ck}_4 {Wk_1})/W_1$\\
$b_5$ & $1.02 \times 10^{-2}$ & y$^{-1}$ & $\big((1-{Ck}_7){Ck}_4 {Wk_1}\big)/W_2 $\\
$b_6$ & $1.12 \times 10^{2}$ & y$^{-1}$ & $ (1-{Ck}_7){Ck}_5R_{CP} $ \\
$b_7$ & $2.21$ & y$^{-1}$ & $(Ck_{6}+Ck_{8} Wk_1)/ W_2$ \\
$b_8$ & $6.58 \times 10^{1}$ & y$^{-1}$ & $ (1-{Ck}_{11}){Ck}_9R_{CP}$\\
$b_9$ & $6.41 \times 10^{-4}$ & y$^{-1}$ & $-Ck_{8} Ck_{11} Wk_1 / W_3 + Ck_{8} Wk_1 / W_3$ \\
$b_{10}$ & $8.21 \times 10^{-1}$ & y$^{-1}$ & ${Ck}_{10} $\\
$b_{11}$ & $3.9 \times 10^{-6}$ & y$^{-1}$ & $Ck_{11} Ck_{8} Wk_1 / W_4$ \\
$b_{12}$ & $4.01 \times 10^{-1}$ & y$^{-1}$ & $({Ck}_{11}{Ck}_9R_{CP}W_3)/W_4 $\\
$b_{13}$ & $8.81 \times 10^{-3}$ & y$^{-1}$ & ${Ck}_{12}$\\
$b_{32}$ & $7.2 \times 10^{-1}$ &y$^{-1}$ & $Ck_{9}$\\

 \hline
\end{tabular}
\end{center}
\caption{\lab{table3part1}Definition of constants (continued below). }
\end{table}

 \begin{table}[!t]
\begin{center}
\centering
\begin{tabular}{|l|l|l|l|}
\hline
Constant & Value & Units& Definition \\
\hline
$b_{35}$ & $9.59 \times 10^{-7}$ &y$^{-1}$ & $(Ck_{11} Ck_{8} Wk_1) / ( R_{CP} W_3) )$ \\
$b_{41}$ & $3.9 \times 10^{-6}$ & y$^{-1}$ & $Ck_{11} Ck_{8} Wk_1 / W_4 $ \\
$b_{42}$ & $4.01 \times 10^{-1}$ &y$^{-1}$  & $ ({Ck}_{11} {Ck}_9 W_3 R_{CP})/W_4$ \\
$b_{81}$ & $1.05 \times 10^{-1}$ & y$^{-1}$ & $Wk_6 Ck_{8} / W_2$ \\
$b_{82}$ & $6.55 \times 10^{-3}$ & y$^{-1}$ & $(1-Ck_{11}) Wk_6 Ck_{8} / W_3$ \\
$b_{83}$ & $3.99 \times 10^{-5}$ & y$^{-1}$ & $Ck_{11} Wk_6 Ck_{8} / W_4$ \\
$b_{84}$ & $9.8 \times 10^{-6}$ &y$^{-1}$ & $(Ck_{11} Ck_{8} Wk_6) / ( R_{CP} W_3) )$ \\
$b_{85}$ & $3.2 \times 10^{-3}$ & & ${Ck}_{13} $ \\
$b_{86}$ & $3.99 \times 10^{-5}$ & y$^{-1}$ & $Ck_{11} Wk_6 Ck_8 / W_4 $ \\
$m_0$ & $3.2 \times 10^{-3}$ & y$^{-1}$ & $(Wk_5+Wk_6) / W_4$\\
$m_{32}$ & $8.34 \times 10^{-2}$ &y$^{-1}$ & $ (Wk_5+Wk_6) / W_3$ \\
$m_{54}$ & $1.05 \times 10^{-1}$ & y$^{-1}$  & $Wk_6 / W_2 $ \\
$m_{56}$ & $7.59 \times 10^{-2}$ & y$^{-1}$  & $Wk_5/W_3$\\
$m_{58}$ & $ 3.2 \times 10^{-3}$ & y$^{-1}$& $(Wk_5+Wk_6) / W_4 $ \\
$m_{71}$ & $4.06 \times 10^{1}$ & y$^{-1}$ & $\big(({Ck}_1 + {Pk}_2)W_1 + {Pk}_4 {Wk_1}\big)/W_1$\\
$m_{72}$ & $3.73 \times 10^{1}$ & y$^{-1}$ &  ${Ck}_1\big(1 - {Pk}_{5a} - (R_{CP}{Pk}_7 {Ck}_3 /400) \big) $\\
$m_{73}$ & $3.87 \times 10^{-1}$ & y$^{-1}$  & $ {Pk}_{5a}{Ck}_1$ \\
$m_{74}$ & $1.07$ & y$^{-1}$ & $Ck_{5}+Pk_{9}+Pk_{11} Wk_1 / W_2$ \\
$m_{75}$ & $1.05  $ & y$^{-1}$ & ${Ck}_5(1 - {Pk}_{14} - {Pk}_{12a}) $\\
$m_{76}$ & $1.06 \times 10^{-2}$ & y$^{-1}$ & ${Pk}_{12a}{Ck}_5$ \\
$m_{77}$ & $6.13 \times 10^{-1}$ &y$^{-1}$ & ${Ck}_9(1 - {Pk}_{20} - {Ck}_{11}) $\\
$m_{78}$ & $7.19 \times 10^{-3}$ & y$^{-1}$ & ${Pk}_{20} {Ck}_9 $ \\
$k_{\text{redox}}$ & $1.00 \times 10^{8}$ & mM y$^{-1}$ & \\
$k_{\text{prec}}$ & $1 \times 10^{-3}$ & mM y$^{-1}$ & \\
$C_{RP}$ & 0.03 &   & [RS]$_{0}$ \\
$g_0$ & $0.17$ & mM  & [O$_2$]$_{t=0}$  \\
$g_s$ & $0.325$ & mM  & $k_{\text{O2--surf}}$  \\
$R_{CP}$ & 106 &   &  \\
$K_m$ & $ 1 \times 10^{-4} $ & mM&\\
$v_a$ & $0.1$ & - &  \\
\hline
\end{tabular}
\end{center}
\caption{\lab{table3part2}Definition of constants continued. In the right hand column, $W_i$ is the volume of box $i$ and ${Wk}_i$, ${Ck}_i$ and ${Pk}_i$ are rate constants in the water, carbon and phosphorus cycles respectively.}
\end{table}

\clearpage

\subsection*{Anoxia parameter computation}

Although our analysis of the model provided us with an extremely simple result (the oxygen status of the deep ocean depending on the sign of $\la_6 s_3-\nu$), the determination of the critical parameters involved in the transition to anoxia is convoluted in the extreme. Therefore here we provide a path to compute them, and in the supplementary material we provide a Matlab code to compute them directly (see Online Resource 1), given the original input parameters of the Slomp model (those listed in tables \ref{table4} and \ref{table5}).

From these, tables \ref{table3part1} and \ref{table3part2} provide definitions of all dimensional parameters. From these, \eqref{B4} provides sequential definitions for all the scales $[r]$, $[s_1]$, etc., where additionally \eqref{2.8.1} and \eqref{2.8.2} have been used; $d_{16}$, $d_{17}$ and $d_{18}$ are defined in \eqref{A2}. From these, \eqref{B1} defines $\la_1$,\ldots$\la_5$, $\la_{11}$ and $\la_{20}$, \eqref{B2} defines $\de_1$,\ldots$\de_5$, \eqref{B3} defines $\eps_1$,\ldots$\eps_4$, $\eps_{6}$, $\eps_{8}$,\ldots$\eps_{11}$, $\eps_{13}$,\ldots$\eps_{16}$, $\eps_{19}$,\ldots$\eps_{39}$, $\eps_{99}$, $\eps_{101}\ldots\eps_{107}$ and $\eps_{110}\ldots\eps_{113}$. We then use \eqref{2.11} to define $\bar{s}_2$ and $\bar{s}_3$, after which \eqref{B5} defines $\la_6$, $\de_{25}$, $\eps_{40}$,\ldots$\eps_{52}$ and $\eps_{120}$,\ldots$\eps_{127}$. Finally, we recover $\la_9$ from \eqref{2.16.2}, $\eps_{57}$ from \eqref{eps57defn} and \eqref{2.18} gives $\eps_{55}$, $\eps_{56}$, $\de_6,\ldots\de_9$ and $\la_{13},\ldots\la_{15}$.

The net result of these transformations is that, in dimensional terms, $\la_6 s_3^{\mbox{\tiny approx}}-\nu$ can be expressed as
\be \lab{lam6s3expr}  \frac{B_5+B_1 B_2}{B_3 B_4 },   \ee
where we have introduced the quantities
\begin{eqnarray*}
 B_1&=& \frac{ a_{36} a_{53} a_{38} a_{27} m_{72} b_{1} a_{45} b_{12} m_{58} m_{74} a_{26} R_{CP} a_{29} a_{43} }{ v_a m_{0} g_s m_{75} b_{13} }, \\ B_2&=& \frac{  \nu (b_{84} b_{32} b_{6}-a_{82} b_{7} m_{77}) v_a-a_{31} b_{7} m_{77}+b_{32} b_{35} b_{6}  }{ v_a m_{0} g_s m_{75} b_{13} }, \\ B_3 &=& X_3 a_{27}+(m_{58} a_{30} m_{77}^2 a_{38} a_{43} a_{45} (a_{81} m_{74} \nu v_a-a_{22} m_{75}) R_{CP}+X_4 a_{36}) b_{7} a_{29}, \\ B_4 &=& m_{71} a_{18}-a_{14} m_{72}, \\ B_5 &=& \frac{ b_{7} a_{29} X_2 m_{75}^2+X_1 m_{75} }{ v_a m_{0} g_s m_{75} b_{13} },
\end{eqnarray*}
which depend on
\begin{eqnarray*}
 X_1&=&-a_{36} (m_{77} ((-m_{54} a_{82}+a_{74} (m_{32}-m_{56})) a_{27}+m_{74} a_{81} (m_{32}-m_{56})) b_{7} \\
 && +b_{84} m_{54} a_{27} b_{32} b_{6}) b_{13} a_{38} \nu^3 a_{45} (a_{14} m_{72}-m_{71} a_{18}) g_s m_{58} R_{CP} m_{0} a_{29} a_{43} v_a^3 \\
 && -b_{13} \nu^2 R_1 (a_{14} m_{72}-m_{71} a_{18}) g_s m_{0} v_a^2-\nu R_2 v_a \\
 &&-a_{36} ((m_{74} b_{32} a_{34} m_{72} a_{26}+a_{27} a_{31} a_{18} a_{21} m_{77}) b_{7} \\
 && -a_{27} b_{32} b_{35} a_{18} a_{21} b_{6}) a_{53} a_{38} b_{1} a_{45} b_{12} m_{58} R_{CP} a_{29} a_{43},\\
 && \\
 X_2&=&\nu^3 a_{83} m_{54} m_{58} b_{32} m_{0} b_{13} a_{36} a_{38} a_{43} a_{45} R_{CP} g_s (a_{14} m_{72}-m_{71} a_{18}) v_a^3 \\
&& +a_{36} b_{13} \nu^2 (a_{14} m_{72}-m_{71} a_{18}) g_s (((a_{22} (m_{32}-m_{56}) m_{77}+a_{34} b_{32} m_{54}) a_{45} m_{58} a_{38} \\
&& +a_{44} a_{40} m_{78} b_{32} m_{54} a_{83}) a_{43} R_{CP}+a_{83} m_{54} b_{32} a_{38} a_{39} b_{42} a_{45}) m_{0} v_a^2 \\
&& -\nu ((-(a_{14} m_{72}-m_{71} a_{18}) g_s m_{0} ((m_{58} b_{32} a_{22} m_{77} a_{45} a_{38}\\
&&+m_{78} a_{40} a_{44} (m_{54} b_{32} a_{34}-m_{56} a_{22} m_{77})) a_{43} R_{CP}\\
&&+a_{38} a_{39} b_{42} a_{45} (m_{54} b_{32} a_{34}-m_{56} a_{22} m_{77})) b_{13}\\
&&+a_{83} m_{58} b_{32} b_{12} a_{18} a_{21} a_{38} b_{1} a_{43} a_{45} a_{53} R_{CP}) a_{36}\\
&&+m_{58} m_{0} b_{13} a_{22} a_{30} m_{77}^2 a_{38} a_{43} a_{45} R_{CP} g_s (a_{14} m_{72}-m_{71} a_{18})) v_a\\
&&-m_{58} b_{32} a_{34} R_{CP} b_{1} b_{12} a_{18} a_{21} a_{36} a_{38} a_{43} a_{45} a_{53},\\
 && \\
 X_3&=&-a_{43} (((m_{54} b_{32} \nu v_a b_{6} (b_{84} \nu v_a+b_{35}) a_{29} \\
 && +(\nu (m_{56}-m_{32}) v_a-b_{32}) a_{23} m_{76} b_{7} m_{77}) a_{36} \\
 && -a_{30} b_{7} m_{77}^2 ((a_{74} \nu v_a+m_{74}) a_{29}-a_{23} m_{76})) m_{58} a_{45} a_{38} \\
 && +a_{44} a_{36} (m_{54} b_{32} b_{6} (b_{84} \nu v_a+b_{35}) a_{29} \\
 && +m_{56} a_{23} m_{76} m_{77} b_{7}) m_{78} a_{40}) R_{CP}\\
 &&-a_{36} (m_{54} b_{32} b_{6} (b_{84} \nu v_a+b_{35}) a_{29}+m_{56} a_{23} m_{76} m_{77} b_{7}) a_{39} b_{42} a_{45} a_{38}, \\
 && \\
 X_4&=&a_{43} (((\nu^2 ((m_{54} a_{82}+a_{74} (m_{56}-m_{32})) a_{27}+m_{74} a_{81} (m_{56}-m_{32})) v_a^2 \\
&& +  \nu ((a_{31} m_{54}-a_{74} b_{32}+m_{74} (m_{56}-m_{32})) a_{27}\\
&&-m_{74} b_{32} a_{81}-a_{22} m_{75} (m_{56}-m_{32})) v_a \\
&& -b_{32} (m_{74} a_{27}-a_{22} m_{75})) m_{77}+\nu m_{54} b_{32} m_{75} v_a (a_{83} \nu v_a+a_{34})) m_{58} a_{45} a_{38} \\
&& + a_{40} ((\nu ((a_{74} m_{56}+a_{82} m_{54}) a_{27}+m_{56} m_{74} a_{81}) v_a \\
&&+(m_{54} a_{31}+m_{56} m_{74}) a_{27}-m_{56} a_{22} m_{75}) m_{77} \\
&& +m_{54} b_{32} m_{75} (a_{83} \nu v_a+a_{34})) a_{44} m_{78}) R_{CP} \\
&&+((\nu ((a_{74} m_{56}+a_{82} m_{54}) a_{27}+m_{56} m_{74} a_{81}) v_a \\
&& +(m_{54} a_{31}+m_{56} m_{74}) a_{27}-m_{56} a_{22} m_{75}) m_{77} \\
&&+m_{54} b_{32} m_{75} (a_{83} \nu v_a+a_{34})) a_{38} a_{39} b_{42} a_{45}.\\
\end{eqnarray*}
Finally, $X_1$ in turn depends on
\begin{eqnarray*}
 R_1&=&((m_{77} ((a_{45} m_{58} ((a_{74} b_{32}-a_{31} m_{54}+m_{74} (m_{32}-m_{56})) a_{27}+m_{74} b_{32} a_{81}) a_{38}\\
 &&-a_{44} ((a_{74} m_{56}+a_{82} m_{54}) a_{27}+m_{56} m_{74} a_{81}) m_{78} a_{40}) a_{43} R_{CP} \\
 && -a_{38} a_{39} b_{42} a_{45} ((a_{74} m_{56}+a_{82} m_{54}) a_{27}+m_{56} m_{74} a_{81})) b_{7}\\
 &&+b_{6} a_{27} b_{32} m_{54} (a_{43} (b_{84} m_{78} a_{40} a_{44}+m_{58} b_{35} a_{38} a_{45}) R_{CP}+b_{84} a_{38} a_{39} b_{42} a_{45})) a_{29}\\
 &&-m_{58} a_{27} b_{7} R_{CP} a_{23} m_{76} m_{77} a_{38} a_{43} a_{45} (m_{32}-m_{56})) a_{36}\\
 &&-m_{58} b_{7} a_{29} a_{30} m_{77}^2 a_{38} a_{43} a_{45} R_{CP} (a_{74} a_{27}+a_{81} m_{74}),\\
 && \\
 R_2&=&(-a_{27} ((((-m_{58} m_{74} b_{32} a_{45} a_{38}+m_{78} a_{40} a_{44} (m_{54} a_{31}+m_{56} m_{74})) a_{43} R_{CP} \\
 && +a_{38} a_{39} b_{42} a_{45} (m_{54} a_{31}+m_{56} m_{74})) m_{77} b_{7} \\
 &&-b_{6} b_{35} b_{32} m_{54} (R_{CP} m_{78} a_{40} a_{43} a_{44}+a_{38} a_{39} b_{42} a_{45})) a_{29} \\
 && -a_{23} (a_{43} (m_{56} m_{78} a_{40} a_{44}-m_{58} b_{32} a_{38} a_{45}) R_{CP}\\
 &&+m_{56} a_{38} a_{39} b_{42} a_{45}) m_{76} b_{7} m_{77}) (a_{14} m_{72}-m_{71} a_{18}) g_s m_{0} b_{13}\\
 &&+a_{53} a_{38} b_{1} a_{45} b_{12} m_{58} R_{CP} ((a_{82} a_{27} a_{18} a_{21} m_{77}+a_{83} m_{74} b_{32} m_{72} a_{26}) b_{7}\\
 &&-b_{84} a_{27} b_{32} a_{18} a_{21} b_{6}) a_{29} a_{43}) a_{36}\\
 &&-m_{0} m_{58} a_{27} b_{7} R_{CP} b_{13} a_{30} m_{77}^2 a_{38} a_{43} a_{45} g_s (a_{14} m_{72}\\
 &&-m_{71} a_{18}) (m_{74} a_{29}-a_{23} m_{76}).
\end{eqnarray*}
Thus, our efforts to write an explicit formula for $\la_6 s_3^{\mbox{\tiny approx}}-\nu$ lead to extremely complicated formulae having no apparent simplification; it thus appears that the simple controlling parameters of the solution depend in a very complicated way on many of the physically prescribed parameters, and this dependence needs to be elucidated computationally (see Online Resource 1).

\subsection*{References}

\begin{description}

\item Abelson, P.\,H.\ 1999 A potential phosphate crisis. Science {\bf  283} (5,410), 2015.

\item Allison, S.\,D.\ and J.\,B.\ Martiny 2008 Resistance, resilience, and redundancy in microbial communities. PNAS {\bf 105} (1), 11512--11519.

\item Anderson, L.\,A.\ and J.\,L.\ Sarmiento 1995 Global ocean phosphate and oxygen simulations. Global Biogeochem.\ Cycles {\bf 9} (4), 621--636.

\item Beil, S.\, W.\  Kuhnt, A.\ Holbourn, F.\ Scholz, J.\ Oxmann, K.\ Wallmann, J.\ Lorenzen, M.\ Aquit and E.\,H.\ Chellai 2020 Cretaceous oceanic anoxic events prolonged by phosphorus cycle feedbacks. Climate of the Past {\bf 16} (2), 757--782.

\item Bergman, N.\,M., T.\,M.\ Lenton and A.\,J.\ Watson 2004 COPSE: a new model of biogeochemical cycling over Phanerozoic time. Amer.\ J.\ Sci.\ {\bf 304}, 397--437.

\item Cordell, D., J.-O.\ Drangert and S.\ White 2009 The story of phosphorus: global food security and food for thought. Global Environmental Change {\bf 19}, 292--305.

\item Filipelli, G.\ ,M.\ 2002 The global phosphorus cycle. In:  Kohn, M., J.\ Rakovan and J.\  Hughes (eds.) Phosphates: geochemical, geobiological, and materials importance. Revs.\   Mineral.\  Geochem.\  {\bf 48}, pp.\ 391--425.

\item Filipelli, G.\ ,M.\ 2008 The global phosphorus cycle: past, present, and future. Elements {\bf 4},  89--95.

\item F\"ollmi, K.\,B.\ 1996 The phosphorus cycle, phosphogenesis and marine phosphate-rich deposits. Earth-Sci.\ Revs.\ {\bf 40}, 55--124.

\item Goldblatt, C., T.\,M.\ Lenton and A.\,J.\ Watson 2006 Bistability of atmospheric oxygen and the Great Oxidation. Nature {\bf 443} (7112), 683--686.

\item Handoh, I.\,C.\ and T.\,M.\ Lenton 2003 Periodic mid-Cretaceous oceanic anoxic events linked by oscillations of the phosphorus and oxygen biogeochemical cycles. Global Biogeochem.\ Cycles {\bf 17} (4), 1092.

\item Jarvis, I., G.\,A.\ Carson, M.\,K.\,E.\ Cooper, M.\, B.\ Hart, P.\,N.\ Leary, B.\,A.\ Tocher, D.\ Horne and A.\ Rosenfeld 1988 Microfossil assemblages and the Cenomanian-Turonian (late Cretaceous) oceanic anoxic event. Cretaceous Res.\ {\bf 9}, 3--103.

\item Jenkyns, H.\,C.\ 2010 Geochemistry of oceanic anoxic events.  Geochemistry Geophysics Geosystems {\bf 11} (3), Q03004.

\item Lenton, T.\,M.\ and A.\,J.\ Watson 2000 Redfield revisited: 2. What regulates the oxygen content of the atmosphere? Global Biogeochem.\ Cycles {\bf 14}, 249--268.

\item Mackenzie, F.\,T., L.\,M.\ Vera and A.\ Lerman 2002 Century-scale nitrogen and phosphorus controls of the carbon cycle. Chem.\ Geol.\ {\bf 190}, 13--32.

\item Maeda, E.\,E., P.\ Haapasaari, I.\ Helle, A.\ Lehikoinen, A.\ Voinov and S.\ Kuikka 2021 Black Boxes and the Role of Modeling in Environmental Policy Making. Front.\ Environ.\ Sci.\ {\bf 9}, 63.

\item Niemeyer, D., T.\,P.\ Kemena, K.\,J.\ Meissner and A.\ Oschlies 2017 A model study of warming-induced phosphorus--oxygen feedbacks in open-ocean oxygen minimum zones on millennial timescales. Earth System Dynamics {\bf 8} (2), 357--367.

\item Ozaki, K., S.\ Tajima and E.\ Tajika 2011 Conditions required for oceanic anoxia/euxinia: constraints from a one-dimensional ocean biogeochemical cycle model. Earth Planet.\ Sci.\ Letts.\ {\bf 304}, 270--279.

\item Percival, L.\,M.\,E., A.\,S.\ Cohen, M.\,K.\ Davies, A.\,J.\ Dickson, S.\,P.\ Hesselbo, H.\,C.\ Jenkyns, M.\,J.\ Leng, T.\,A.\ Mather, M.\,S.\ Storm and W.\ Xu 2016 Osmium isotope evidence for two pulses of increased continental weathering linked to Early Jurassic volcanism and climate change. Geology {\bf 44} (9), 759--762.

\item  Percival, L.\,M.\,E., M.\,L.\,I.\ Witt, T.\,A.\ Mather, M.\ Hermoso, H.\,C.\ Jenkyns, S.\,P.\ Hesselbo, A.\,H.\ Al-Suwaidi,  M.\,S.\ Storm, W.\ Xu and  M.\ Ruhl 2015 Globally enhanced mercury deposition during the end-Pliensbachian extinction and Toarcian OAE: a link to the Karoo?Ferrar Large Igneous Province. Earth Planet.\ Sci.\ Letts.\ {\bf 428}, 267--280.

\item Pianosi, F., K.\ Beven, J.\ Freer, J.\,W.\ Hall, J.\ Rougier, D.\,B.\ Stephenson and T.\ Wagener 2016 Sensitivity analysis of environmental models: A systematic review with practical workflow. Environ Model Softw {\bf 79}, 214--232.
\item Schlanger, S.\,O.\ and H.\,C.\ Jenkyns 1976 Cretaceous oceanic anoxic events: causes and consequences. Geologie en Mijnbouw {\bf 55} (3-4), 179--184.

\item Sell, B., M.\ Ovtcharova, J.\ Guex, A.\ Bartolini, F.\ Jourdan, J.\,E.\ Spangenberg, J.-C.\ Vicente  and U.\ Schaltegger 2014 Evaluating the temporal link between the Karoo LIP and climatic-biologic events of the Toarcian Stage with high-precision U-Pb geochronology. Earth Planet.\ Sci.\ Letts.\ {\bf 408}, 48--56.

\item Slomp, C.\,P.\ and P.\ van Capellen 2007 The global marine phosphorus cycle: sensitivity to oceanic circulation. Biogeosci.\ {\bf 4}, 155--171.

\item Tsandev, I., C.\,P.\ Slomp and P.\ van Capellen 2008 Glacial-interglacial variations in marine phosphorus cycling: implications for ocean productivity. Global Biogeochem.\ Cycles {\bf 22}, GB4004.

\item Turgeon, S.\,C.\ and R.\,A.\ Creaser 2008 Cretaceous oceanic anoxic event 2 triggered by a massive magmatic episode. Nature {\bf 454}, 323--326.

\item Tyrrell, T.\ 1999 The relative influence of nitrogen and phosphorus on oceanic primary production. Nature {\bf 400}, 525--531.

\item Van Cappellen, P.\ and E.\,D.\ Ingall 1994 Benthic phosphorus regeneration, net primary production, and ocean anoxia: a model of the coupled marine biogeochemical cycles of carbon and phosphorus. Paleoceanography {\bf 9} (5), 677--692.

\item Van Cappellen, P.\ and E.\,D.\ Ingall 1996 Redox stabilization of the atmosphere and oceans by phosphorus-limited marine
productivity. Science {\bf  271} (5,248),  493--496.

\item Wallmann, K., S.\, Fl{\"o}gel, F.\, Scholz, A.\,W.\, Dale, T.\,P.\, Kemena, S.\ Steinig and W.\ Kuhnt 2019 Periodic changes in the Cretaceous ocean and climate caused by marine redox see-saw. Nature Geoscience {\bf 12} (6), 456--461.

\item Watson, A.\,J.\ 2016  Oceans on the edge of anoxia. Science {\bf 354} (6319), 1529--1530.

\end{description}

\end{document}